\documentclass[11pt]{article} 
\usepackage[utf8]{inputenc} 

\usepackage{geometry} 
\usepackage{hyperref}
\usepackage{amsmath}
\usepackage{graphicx} 

\usepackage{amsmath, amssymb, amsthm}
\usepackage{enumerate}
\usepackage{bm}
\usepackage{mathrsfs}
\usepackage[square,numbers]{natbib}
\usepackage{booktabs} 
\usepackage{array} 
\usepackage{paralist} 
\usepackage{verbatim} 
\usepackage{subfig} 

\usepackage{fancyhdr} 
\usepackage{sectsty}
\allsectionsfont{\sffamily\mdseries\upshape} 

\usepackage[nottoc,notlof,notlot]{tocbibind} 
\usepackage[titles,subfigure]{tocloft} 

\newcommand{\be}{\begin{equation}}
\newcommand{\ee}{\end{equation}} 
\newcommand{\bs}{\boldsymbol}

\title{Optimal Diversification and Leverage in a Utility-Based Portfolio Allocation Approach}
\author{Vladimir Markov}
\date{} 
\begin{document}
\maketitle

\begin{abstract}
We examine the problem of optimal portfolio allocation within the framework of utility theory. We apply exponential utility to derive the optimal diversification strategy and logarithmic utility to determine the optimal leverage. We enhance existing methodologies by incorporating compound probability distributions to model the effects of both statistical and non-stationary uncertainties. Additionally, we extend the maximum expected utility objective by including the variance of utility in the objective function, which we term generalized mean-variance. In the case of logarithmic utility, it provides a natural explanation for the half-Kelly criterion, a concept widely used by practitioners.
\end{abstract}

\section{Introduction}
Portfolio allocation is a classical problem in finance \cite{Markowitz:1952}, with many models proposed over time. Despite the multitude of available approaches, at its core, a trader needs prescriptions for two key parameters: diversification across securities and the leverage, which dictates the percentage of risk capital allocated in each sequential bet.

Managing uncertainty is essential both across securities and over time in sequential decision-making, as these sources of uncertainty are intertwined. In any realistic scenario, the problem of portfolio optimization becomes unsolvable due to the rapidly accumulating  uncertainty at each time step and the presence of exogenous information flow. While the former can, in principle, be estimated, the latter is inherently unpredictable. Consequently, we propose to optimize only the next time slot while accounting for the risks ahead. After executing each allocation, we observe the outcome, update the data, and re-optimize.

The cross-sectional allocation for the next timeslot is determined using a single-period optimization approach, where risk is mitigated through diversification. For this purpose, we employ the exponential (CARA) utility function \cite{Arrow:1966, Pratt:1964}. Over multiple periods, traders face the risk of ruin inherent in multiplicative processes like geometric Brownian motion (GBM). Even with a positive mean, GBM causes most trajectories to trend toward zero wealth. To address this risk, we utilize the logarithmic utility function \cite{Kelly:1956}, applied to the final wealth evolving under GBM with parameters determined by maximizing the exponential utility. We incorporate the variance of logarithmic utility into our objective to mitigate the risk from short-term fluctuations in wealth and derive an analytical solution for the optimal leverage.

In this paper, we propose a few enhancements to the existing methodologies. By accounting for the variance of utility, we extend the maximum expected utility objective into a generalized mean-variance framework. We also adopt a compound distribution for the investment outcome. This approach introduces hyperparameters that allow us to model both statistical noise and non-stationary effects in the expected returns and covariance matrices. We argue that incorporating this intermediate statistical layer between the utility formulation and the optimization process offers a flexible framework for making decisions under uncertainty.

\section{The generalized mean-variance and compound distribution of the outcome}
 
Suppose the outcome of interest is a random variable $X$, and the utility of the outcome is $Y = U(X)$. The outcome $X$ is a function of decision variables, which in our case are portfolio weights $w$ (to be defined). The utility $U(X)$ is a non-linear transformation of $X$  and encodes the investor's preferences towards profitability and risk (volatility of the outcome). 

A practical framework for modeling uncertainties is the compound probability distribution formalism.
In this approach, we assume that the location $\mu$ and scale $\sigma$ parameters of the outcome distribution $X\sim P(\mu,\sigma)$ are parameterized by distributions $G_{\mu}$ and $G_{\sigma}$ with corresponding hyperparameters $\theta_{\mu}$ and $\theta_{\sigma}$:
\be 
X_t \sim P(\mu, \sigma), \,\,
\mu \sim G_{\mu}(\theta_{\mu}|D_{t-1}), \,\,
\sigma^2 \sim G_{\sigma}(\theta_{\sigma}|D_{t-1})
\label{eq:compound1}
\ee
where the hyperparameters $\theta_{\mu}$ and $\theta_{\sigma}$ are conditioned on past data and subjective views on the market, $D_{t-1}$. 

The standard approach to determining decision variables is maximizing the expected utility $\text{E}[U(X)]$. For a non-linear utility, volatility risk is accounted for by using the second moment of the Taylor expansion of $U(X)$. In this paper, we investigate an additional component of risk, parameterized by the variance of the utility  $ \text{Var}[U(X)] $, which encodes the decision maker's aversion to dispersion around the expected utility. This introduces a hyper-utility function that measures the trade-off between maximizing expected utility and minimizing the dispersion around its mean which we call a generalized mean-variance (GMV) problem: 
\be
w = \arg\max_w \left( E[U(X_t|D_{t-1})] - \frac{\lambda}{2} \, \text{Var}[U(X_t|D_{t-1})] \right)
\label{eq:gmv1}
\ee
where $U(X_t)$ represents classical utilities, which can be linear, exponential (CARA), power-law (CRRA), or logarithmic,  $w$ are decision variables (allocation weights),  and $\lambda$ is an utility risk aversion parameter.  The solution coincides with the maximum expected utility (MEU) objective for  $\lambda=0$. We employ the GMV of the logarithmic utility to obtain the optimal leverage.

To avoid cluttering notation, we drop time index $t$ below, always assuming that we use the past data  $D_{t-1}$ to predict parameters of the future utility $U(X_t)$.

The expected value of the utility is obtained by integrating over the distribution outcomes $P$ with parameters $\theta \in \{\mu, \sigma^2\}$ and $G$ defined by its corresponding hyperparameters:
\be
E[U(X)] = E_{P}[E_G[U(X|\theta)]] =  \int dX \, U(X) \int d\theta \,  P(X|\theta) G(\theta|D) 
\ee 
Integrating over the parameters $\theta$, as expressed by 
$\int d\theta \, P(X|\theta) G(\theta|D),$  is formally equivalent to computing the predictive distribution (PD) of the outcome $X$ in Bayesian statistics. For instance, if $G(\theta|D)$ is the posterior distribution of $\theta$, the result is the posterior predictive distribution (PPD); if $G(\theta)$ is the prior distribution, it is the prior predictive distribution.
The PPD incorporates information about the noise in the outcome $X$, the uncertainty of the parameters $\theta$, given a finite sample size of the data $D$ and priors. 
The variance of $U(X)$ is given by
\be 
\text{Var}[U(X)] =  E[(U(X) - E[U(X)])^2]  =E[U(X)^2] - E[U(X)]^2 
\ee
In Appendix A, we discuss general aspects of the GMV approach, and in Appendix B, we derive closed-form formulas for the GMV for the most common choices: linear, logarithmic, exponential (CARA), and power (CRRA) utilities.

\section{Optimal diversification} 
To reduce the risk from uncertainty about the future, diversification is needed.
If there is certainty or complete conviction about the future performance of assets, the optimal allocation is to invest all capital in the best-performing asset. Conversely, in cases of complete uncertainty, a maximum entropy equal-weight allocation is optimal. In the real world, the future expected returns and the portfolio covariance matrices have limited predictability. 
This situation requires finding an optimal diversification strategy based on available information, which interpolates between these two regimes. 

We use the exponential utility $U_a(x,a)$ with respect to the investment outcome $x$ and the risk aversion parameter $a$:
\be
 U_a(x,a) = 
    \begin{cases}
       \frac{1-\exp(-a x)}{a}, &  \mbox{if }a \neq 0 \\
        x, &  \mbox{if }  a = 0
    \end{cases}, \,\, \,
\label{eq:utility}
\ee
which encodes a trader's risk preferences, and use compound distributions to account for uncertainties as the main tools for deriving optimal diversification. The outcome can be expressed as $x = W_0(1+(1-\boldsymbol w^T \boldsymbol I) r_0 + \bs w^T \bs r)$, where $W_0$ is the initial wealth, which we always rescale to 1, $r_0$ is the risk-free rate, $\bs w$ is the allocation to a risky asset, and $\bs r$ is the return of the risky asset.

This paper relies on analytical results from \cite{MarkovA:2023,MarkovB:2023}. The application of exponential utility to portfolio optimization was also studied in \cite{Birge:2020,Boyd:2022,Giller:2004}. The Bayesian treatment of the MV approach was presented in \cite{Bodnar:2020}. Although the MV approach can be seen as a particular case of optimizing the expected value of exponential utility under Gaussian returns, our results differ due to the non-commutativity of the expected value over uncertain parameters and $\arg\max$ operations. The main thesis of this paper is that, to account for realistic features of portfolio optimization—fat tails, return skewness, and uncertainty in the future mean and (co)variance—the starting point should be the expected-utility formulation rather than mean–variance.

In this section, we concentrate on deriving analytical results for the application of exponential utility with an uncertain mean $\bs \mu$ and covariance matrix $\bs \Sigma$. If the future covariance matrix $\boldsymbol \Sigma$ and the expected returns $\boldsymbol \mu$ are random variables, the optimal allocation weights are determined by integrating over all possible realizations of the covariance matrix $\boldsymbol \Sigma$ 
and the expected returns $\boldsymbol \mu$. 
The optimal weights $\bs w^*$ are given by
\be
\bs w^*=\arg \max_{\bs w} E[U_a(\bs \mu,\bs\Sigma)]=\arg \max_w \int d x\,\, \int d \boldsymbol \Sigma \,\, \int d \boldsymbol \mu \,\,   U_a(x,a) P(x;\boldsymbol \mu,\boldsymbol \Sigma) P(\boldsymbol \mu,\boldsymbol \Sigma|D) 
\label{formula-port}
\ee
where $P(x;\boldsymbol \mu,\boldsymbol \Sigma)$ is the distribution of the outcome $x$ given $\boldsymbol \mu$ and $\boldsymbol \Sigma$, and $P(\boldsymbol \mu,\boldsymbol \Sigma \mid D)$ is the posterior distribution of $\boldsymbol \mu$ and $\boldsymbol \Sigma$, given the data $D$, the data-generating process (likelihood), and the priors.

A large number of analytical objective functions can be obtained by observing two facts. 
First, the integral over the parameters $\bs \mu$ and $\bs \Sigma$ formally coincides with the posterior predictive distribution $P_{ppd}(x)$ of the outcome $x$ in Bayesian statistics, provided that $P(\bs \mu,\bs \Sigma|D)$ is the posterior distribution of $\bs \mu$ and $\bs \Sigma$:
\be
P_{ppd}(x|D) \sim \int d \boldsymbol{\Sigma} \, \int d \boldsymbol{\mu} \, P(x; \boldsymbol{\mu}, \boldsymbol{\Sigma})
P(\boldsymbol \mu,\boldsymbol \Sigma|D) 
\ee  
The closed analytical form of the PPD is known for a multivariate normal outcome in the case of an uncertain mean, covariance matrix, or both. If the PPD is not available in closed form or does not integrate analytically with the utility, we use a heuristic noise model for $P(\bs \mu,\bs \Sigma|D)$ to achieve analytical tractability of the problem.
Often assuming the factorization property
$P(\boldsymbol{\mu},\boldsymbol{\Sigma}|D) = P(\boldsymbol{\mu}|D) P(\boldsymbol{\Sigma}|D)$ to simplify the modeling is enough. Eq.~\ref{formula-port} is applicable to an arbitrary utility function with uncertain parameters for the outcome $x$.

Second, the expected value of exponential utility for arbitrary distribution $P(x)$ is provided by its moment generating function (MGF):
\be
E [U_a (X)] = \frac{1}{a}\left(1 - E[e^{-a X}]\right) =\frac{1}{a}(1 - MGF(-a))
\ee
which is readily available for all major statistical distributions. We apply a logarithmic transformation to the expected utility to obtain an interpretable objective function. The maximization of the exponential utility $E(U_a(X))$ also appeared in the literature under different names: as the minimization of the entropic risk measure defined as $\rho^{\mathrm{ent}}(X)=\frac{1}{a}\ln E[e^{-a X}]$, or as the maximization of the certainty equivalent of the CARA utility $\mathrm{CE}(X)=-\frac{1}{a}\ln E[e^{-a X}]$.

For normally distributed returns $\bs r \sim N(\bs \mu, \bs \Sigma)$ with an uncertain location vector $\boldsymbol\mu \sim \mathcal{N}(\boldsymbol \mu_0, \boldsymbol \Sigma_0)$, the optimal weights $\boldsymbol w^*$ after marginalization over $\bs \mu$ are given by \cite{MarkovA:2023}
\be
\boldsymbol w^*=\arg \max_{\boldsymbol w} \left[ (1-\boldsymbol w^T \boldsymbol I) r_0+ \boldsymbol \mu_0^T \boldsymbol w-\frac{a}{2}  \boldsymbol w^T \boldsymbol  (\bs \Sigma_0+\bs \Sigma) \boldsymbol w \right]
\label{eq:opt-weight3aB0}
\ee
Here $\bs \Sigma_0$ is a diagonal matrix of variances of the expected return $\sigma_0^2$, $\mathbf{\Sigma}$ is a covariance matrix.  

For normally distributed returns $\bs r \sim \mathcal{N}(\bs \mu, \bs S)$ with an uncertain location vector $\boldsymbol\mu \sim \mathcal{N}(\boldsymbol \mu_0, \boldsymbol \Sigma_0)$ and an uncertain Wishart-distributed covariance matrix $\bs S \sim W_N(\alpha, \frac{\bs \Sigma}{\alpha})$, the optimal weights $\boldsymbol w^*$ after marginalization over $\bs \mu$ and $\bs S$ are given by \cite{MarkovB:2023}
\be
\boldsymbol w^*=\arg \max_{\boldsymbol w} \left[ (1-\boldsymbol w^T \boldsymbol I) r_0+ \boldsymbol \mu_0^T \boldsymbol w-\frac{a}{2}  \boldsymbol w^T \boldsymbol  \Sigma_0 \boldsymbol w +\frac{\alpha}{2 a} \ln \left[1-\frac{a^2}{\alpha}  (\boldsymbol w^T \boldsymbol  \Sigma \boldsymbol w)\right] \right],
\label{eq:opt-weight3aB}
\ee
such that $1-\frac{a^2}{\alpha}  (\boldsymbol w^T \boldsymbol  \Sigma \boldsymbol w)>0$. Here $\alpha$ is a parameter that controls the noise level of the covariance matrix. 

For multivariate asymmetric Laplace distributed (ALD) returns $\bs r \sim ALD(\bs \mu,\bs \Sigma,\bs \mu_a)$ with an uncertain location vector $\boldsymbol\mu\sim \mathcal{N} (\boldsymbol \mu_0,\boldsymbol \Sigma_0)$, we have the following optimization objective \cite{MarkovA:2023}:
\be
\boldsymbol w^*=\arg \max_{\boldsymbol w} \left[  (1-\boldsymbol w^T \boldsymbol I) r_0+ \boldsymbol \mu_0^T \boldsymbol w-\frac{a}{2}  \boldsymbol w^T \boldsymbol \Sigma_0 \boldsymbol w+\frac{1}{a} \ln\left[1-\frac{a^2}{2} \boldsymbol  w^T \boldsymbol \Sigma \boldsymbol w+a  \boldsymbol \mu_a^T \boldsymbol w\right] \right],
\label{eq:opt-weight4aB}
\ee
such that $1-\frac{a^2}{2} \boldsymbol  w^T \boldsymbol \Sigma \boldsymbol w+a  \boldsymbol \mu_a^T \boldsymbol w>0$. Here $\bs \Sigma_0$ is a diagonal matrix of variances of the expected return $\sigma_0^2$, $\mathbf{\Sigma}$ is a scale matrix , and $\boldsymbol{\mu_a}$ is an asymmetry parameter.

The objectives are convex functions of the weights $\bs w$  and can be solved numerically using a convex optimizer, with additional constraints incorporated if necessary. In Appendix C, we discuss the analytical solution of the equations for the case $\boldsymbol \Sigma_0 = 0$ and no constraints.

\subsection{Effect of statistical noise and non-stationarity}

The interpretation of hyperparameters differs between stationary and non-stationary settings.  In the stationary setting, statistical noise arises from finite-sample estimates and observational noise. In the latter, additional noise originates from the changing mean and variance of the generative process.

In the stationary setting, we model univariate returns as $ r \sim \mathcal{N}(\mu,\sigma^2)$ with $\mu \sim \mathcal{N}( \mu_{pd},\sigma_{pd}^2)$, where $\mu_{pd}$ and $\sigma_{pd}^2$ 
are the posterior mean and variance, respectively.  After marginalization over the mean $\mu$, we have $ p(r|D) = \int p(r | \mu) p(\mu) \, d\mu$. $p(r|D)$ coincides with the PPD of a normal random variable with an uncertain mean: $p(r|D)= \mathcal{N}(\mu_{pd}, \sigma^2+\sigma_{pd}^2  )$.
The variance of the PPD is the uncertainty due to the observation noise $\sigma^2$, plus the uncertainty due to the parameter estimation $\sigma_{pd}^2$. The posterior mean $\mu_{pd}$ and its posterior variance $\sigma_{pd}^2$ are given by the formulas for the Gaussian posterior distribution with unknown mean and known variance:
\be
\mu_{\text{pd}} = \sigma_{\text{pd}}^2 \left(\frac{\tilde\mu_0}{\tilde\sigma_0^2} +  \frac{n \bar r}{\sigma^2}\right),
\,\,\,
\sigma_{\text{pd}}^{-2} = \frac{1}{\tilde \sigma_0^2} +  \frac{n}{\sigma^2}
\label{eq:bayes_update_1d}
\ee
where $\tilde \mu_0$ and $\tilde \sigma_0^2$ are the prior mean and variance,
$\sigma^2$ is approximated using the observed data variance, $\bar r$ is the mean return, and $n$ is the number of observations. 
For the multivariate case  $ \bs r \sim \mathcal{N}(\bs \mu,\bs \Sigma)$ with the posterior distribution $\bs \mu \sim 
\mathcal{N}(\boldsymbol{{\mu}}_{pd}, \boldsymbol{{\Sigma}}_{pd})$,
the posterior and predictive posterior are given by
\be
p(\bs \mu | D, \bs \Sigma) = \mathcal{N}(\bs \mu | \bs \mu_{pd},\bs \Sigma_{pd}), \,\,\,\, p(\bs r | D) = \mathcal{N}(\bs r | \bs \mu_{pd}, \bs \Sigma + \bs \Sigma_{pd}) 
\ee
where 
\be
\bs \mu_{pd} = \bs \Sigma_{pd} \left( \boldsymbol{\tilde{\Sigma}}_0^{-1} \boldsymbol{\tilde{\mu}}_0+n \bs \Sigma^{-1} \bar{\bs r}  \right), \,\,
\bs \Sigma_{pd}^{-1} =  \boldsymbol{\tilde{\Sigma}}_0^{-1}+ n \bs \Sigma^{-1}\,\,\, 
\label{eq:bayes_update_2d}
\ee
Here $\boldsymbol{\tilde{\mu}}_0$ and $ \boldsymbol{\tilde{\Sigma}}_0^2$ are the prior mean and covariance matrix,
$\bs \Sigma^2$ is approximated using the observed data covariance matrix, $\bar{\bs r}$ is the mean return, and $n$ is the number of observations. 

A heuristic way to account for non-stationarity is to replace the number of observations $n$ in Eq.~\ref{eq:bayes_update_1d} or \ref{eq:bayes_update_2d} with an effective number of observations $n^*$ such that $n^* < n$. 
It is worth noting that the PPD is also known for the univariate and multivariate cases of Gaussian outcomes with Normal-Wishart and Normal-Inverse-Wishart priors. In both cases, the PPD is given by the Student's $t$-distribution, which, in the case of exponential utility, can only be studied numerically.

Finance is inherently an open, non-equilibrium system that continuously exchanges information and capital with its environment.
This openness is one of the main reasons why financial time series are non-stationary. Technically, return non-stationarity can be attributed to the variation of expected return, the covariance matrix over time, and regime changes. The problem of portfolio allocation is most sensitive to the variation of expected return over time. If the expected return changes sign during the holding period $T$ due to non-stationarity, there is no reason to hold an asset, even if it improves diversification.
We model time varying expected return in a univariate setting to gain a conceptual understanding.  Consider a univariate arithmetic Brownian motion (ABM) stochastic differential equation for the price $X_t$:
\be 
    dX_t = \mu_t dt + \sigma dW_t, \, \, \, d\mu_t = \sigma_\mu dB_t,
\label{eq:mu_nonstat}    
\ee
where $W_t$ and $B_t$ are Brownian motions, and the initial condition is given by estimates from the posterior distribution
\begin{equation}
    \mu_{t=0} \sim \mathcal{N}(\mu_{pd}, \sigma_{pd}^2).
\end{equation}
This class of models are well known in  economics \cite{Cochrane:2011}, in  econometrics as local level model and in theory of Kalman filtering \cite{DurbinKoopman:2012}. More complete model should include wandering (co)-variance and correlated noises but lacks analytical solution. As we show in Appendix D, the solution is given by
\be
X_t \sim N\Bigl(X_0 + \mu_{pd} t,\; \sigma^2 t + \sigma_{pd}^2 t^2 +  \frac{\sigma_\mu^2 t^3}{3} \Bigr)
\label{eq:abm_ns}
\ee
Note that drift uncertainty scales as $t^2$ and drift non-stationarity as $t^3$, both dominating the diffusion term in the limit $t\to\infty$, which scales as $t$. For small $t\ll 1$ (volatility is typically measured annually, in which case $t=1$ corresponds to one year), the diffusion contribution dominates the other terms.
The price increment is given by
\be
\Delta X_t=X_{t+\Delta}-X_t \sim N\Bigl(\mu_{pd} \Delta,\; \sigma^2 \Delta + \sigma_{pd}^2 \Delta^2 +  \sigma_\mu^2( t \Delta^2 +\frac{\Delta^3}{3})\Bigr)
\label{eq:abm_ns_inc_t}
\ee
A notable consequence for risk management is that the variance of increment depends on time $t$.                                                                                                                                                                                                                                                                                                                                                                                                                                                                                                                                                                                                                                                                                                                                                                                                                                                                                                                                                                                                                  
The return for a holding period of length \(\Delta = T\), starting at \(t=0\), is given by
\be r_T=\frac{X_T-X_{0}}{X_{0}}\sim N\Bigl(\mu_{pd} T,\; \sigma^2 T + \sigma_{pd}^2 T^2 +   \frac{\sigma_\mu^2 T^3}{3}\Bigr)
\label{eq:abm_ns_ret}
\ee
We use Eq.~\ref{eq:abm_ns_ret} to find the mean and variance of future returns. For example, by integrating the exponential utility function with this distribution, we obtain the mean-variance (MV) allocation for a risk asset:
\be
w^*=\frac{1}{a}  \frac{\mu_{pd}-r_0}{\sigma^2 + \sigma_{pd}^2 T +  \frac{ \sigma_\mu^2 T^2}{3}}
\ee
The posterior distribution of the mean $\mu_{pd}$ and variance $\sigma_{pd}^2$ can be obtained numerically using an MCMC algorithm or through a Kalman filtering procedure for a linear–Gaussian state–space model \cite{ChuiChen:2017}.

We provide analytical solutions for the price distribution with a non-stationary mean for both ABM and geometric Brownian motion (GBM) processes in Appendix D.

Calibrating Eq.~\ref{eq:mu_nonstat} in the multivariate case is technically challenging, so we assume that the matrix of uncertainties in the expected return, $\bs \Sigma_0$, is diagonal, with entries given by the variance of the univariate solution of Eq.~\ref{eq:abm_ns}.


The parameter $\alpha$ in Eq.~\ref{eq:opt-weight3aB} can be used to account for the statistical uncertainty in the estimation of the covariance matrix. Lower values of $\alpha$ correspond to higher uncertainty about future covariance matrix.

\subsection{Compound distributions and marginalization over parameters}
In this section, we discuss three key consequences that result from marginalizing over $\bs \mu$ and $\bs \Sigma$ in Eq.~\ref{formula-port}.

First, for Gaussian returns $\bs r \sim \mathcal{N}(\bs \mu, \bs \Sigma)$ with $\bs \mu \sim \mathcal{N}(\bs \mu_0, \bs \Sigma_0)$, the optimal weights are given by the solution of Eq.~\ref{eq:opt-weight3aB0}:
\be
\bs w^* = \frac{1}{a} (\bs \Sigma + \bs \Sigma_0)^{-1} (\bs \mu_0-r_0 \bs I)
\label{eq:mv_un_mu}
\ee
Uncertainty in expected returns $\bs\Sigma_0$ shifts the covariance matrix from $\bs\Sigma$ to $\bs\Sigma + \bs \Sigma_0$, where $\bs \Sigma$ is the empirical covariance matrix, and $\bs \Sigma_0$ is a diagonal matrix of the variances of the expected returns $\bs \mu$. 
The  instability of mean-variance optimization can be attributed to the small eigenvalues of the covariance matrix $\bs \Sigma$. The eigenvalues of $\bs \Sigma + \bs \Sigma_0$ are shifted upward, making the inversion more stable, similar to the widely used shrinkage approach. 
As we noted before, although  $\bs \Sigma$ scales linearly with the holding period $T$, $\bs \Sigma_0$ scales quadratically $T^2$ and cubically $T^3$. For a long enough horizon $T$, these new terms make a non-negligible contribution to the covariance matrix.


The second point is that the widely used normalization of weights, $\sum_{i=1}^N w_i = 1$, is not required in the exponential utility formulation. If outcomes have non-trivial properties—such as fat tails, skewness of returns, or uncertainty in the expected return and covariance matrix—a scaling factor emerges, which incorporates much of the information about these model assumptions. Normalizing the weights discards this information. 
We derived a closed form for the scaling factor $g$ in the case of the Laplace distribution (fat tails factor) and marginalization over covariance matrices with Wishart-distributed noise (covariance matrix uncertainty factor)  with $\bs \Sigma_0 = 0$ and without constraints. 
For ALD returns,  the optimal weights for  Eq.~\ref{eq:opt-weight4aB} are given by
\be
\boldsymbol{w^*} = g_{ALD}(q,v)\times \left( \frac{1}{a} \boldsymbol{\Sigma}^{-1} (\boldsymbol{\mu} - r_0 \boldsymbol{I})\right) + \frac{1}{a} \boldsymbol{\Sigma}^{-1} \boldsymbol{\mu}_a, \,\, \,\, g_{ALD}(q,v) = \frac{\sqrt{1 + 2 q + q v}-1}{q} 
\ee
where $q = (\boldsymbol{\mu} - r_0 \boldsymbol{I})^T \boldsymbol{\Sigma}^{-1} (\boldsymbol{\mu} - r_0 \boldsymbol{I})$ and $v = \boldsymbol{\mu}_a^T \boldsymbol{\Sigma}^{-1} \boldsymbol{\mu}_a$.
For Gaussian returns with a Wishart-distributed covariance matrix, the optimal weights for Eq.~\ref{eq:opt-weight3aB} are given by
\be 
\bs w^* = g_W(q,\alpha) \times \left(\frac{1}{a} \bs \Sigma^{-1} (\bs \mu-r_0 \bs I)\right),\,\, \,\, \,\, g_W(q,\alpha)= \frac{1}{2 q} \left(\sqrt{\alpha(4 q+\alpha)}-\alpha\right)
\ee
The scaling factor $g_W(q,\alpha)$ is a function of $q$ and decreases with uncertainty $\frac{1}{\alpha}$. The parameter $q$ is the square of the theoretical Sharpe ratio of the MV portfolio. The expected return and risk of the portfolio is given by
\be
\mu_p=\bs w^{*T} (\bs\mu-r_0 \bs I)=g_W(q,\alpha) \frac{q}{a},\,\,\, \sigma_p^2=\bs w^{*T} \bs \Sigma \bs w=g_W(q,\alpha)^2 \frac{q}{a^2}
\label{eq:allocation1}
\ee
The Sharpe ratio is given by $SR=\frac{\mu_p}{\sigma_p}=\sqrt{q}$. 

A similar scaling factor appears in the risk budget formulation of portfolio optimization, maximizing the expected return subject to a risk target constraint $\sigma_{target}$ for the portfolio volatility $\sigma_p \le \sigma_{target}$:
\be
\arg\max_{\bs w} E \left[\bs w^T \bs r\right] = \bs w^T (\bs \mu-r_0 \bs I) 
\quad \text{s.t.} \quad \text{Var} \left[\bs w^T \bs r\right] = \bs w^T \bs \Sigma \bs w \leq \sigma_{target}^2,
\label{eq:risk_budget}
\ee
where $\bs w$ is an allocation weight vector and $\bs r$ is a vector of returns. The solution is given by $w_{rb}^* = \frac{\sigma_{\text{target}}}{\sqrt{q}} \bs \Sigma^{-1} (\bs \mu-r_0 \bs I)$ and Sharpe ratio $SR_{rb}=\sqrt{q}$. 

In case of non-zero $\bs \Sigma_0$, the Eq.~\ref{eq:opt-weight3aB} can be solved by a convex optimizer numerically. 
The expected return of the portfolio $\mu_p$, its variance $\sigma^2_{0p}$ and portfolio risk $\sigma_p^2$ are given by
\be
\mu_p=\bs w^{*T} (\bs \mu_0-r_0\bs I),\, \sigma^2_{0p}=\bs w^{*T} \bs \Sigma_0 \bs w^*,\,  \sigma_p^2=\bs w^{*T} (\bs \Sigma+\bs \Sigma_0) \bs w^*
\label{eq:allocation2}
\ee
For $\bs \Sigma_0>0$, the Sharpe ratio $SR=\frac{\mu_p}{\sigma_p}<\sqrt{q}$ . 
Eq.~\ref{eq:allocation2} is used in the next section as the input to the optimization of the Kelly leverage in the GMV framework. 

Third, the role of the logarithm is the following. The risk term 
$ \ln \left[ 1 - c (\boldsymbol{w}^T \bs \Sigma \boldsymbol{w}) \right] $
resembles the standard MV term for small risk, since for small 
$ x = (\boldsymbol{w}^T \Sigma \boldsymbol{w}) $, we have $ \ln(1 - c x) \approx -c x $. 
For larger risk $x$, a logarithmic risk singularity appears at $x = \frac{1}{c}$ due to the ALD distribution's fat tails or the marginalization of an uncertain covariance matrix for Gaussian returns.

\subsection{Calibration of risk-aversion parameter $a$}
The exponential utility function includes a risk-aversion parameter  $a$, which requires calibration.
To address this problem, we employ a certainty-equivalent (CE) measure of risk.  
The CE is defined as the lowest amount of guaranteed money that a decision-maker would be willing to accept instead of engaging in a gamble or a one-period portfolio trade. 
If empirical data or a subjective view of the market provides $x_c$ as the CE for a gamble with an expected value $\mu_c$ and variance $\sigma_c^2$, the risk aversion parameter  $a$ can be determined by solving \( U(x_c) = E[U(x)] \):
\be
x_c = \mu_c - \frac{a (\sigma_c^2+\sigma_0^2)}{2}.
\ee
where $\mu_c = E[\bs x]= \sum_{i=1}^n p_i x_i$ and $Var[\bs x] = \sigma_c^2=\sum_{i=1}^{n} p_i (x_i - \mu_c)^2$  and $p_i$ is the  probability of the outcome $x_i$
and $\sigma_0^2$ is a variance of the expected return $\mu_c$. Rearranging this equation yields:
\be
a = 2 \frac{\mu_c - x_c}{\sigma_c^2+\sigma_0^2},
\label{eq:risk_a}
\ee
For exponential utility with ALD returns $r_t\sim ALD(\mu,\sigma,\mu_a)$, we have the following equation:
\be 
x_c = \mu_c-\frac{a}{2}\sigma_0^2 +\frac{1}{a} \ln \left[1-\frac{a^2}{2} \sigma_c^2+a \mu_a\right]
\ee
For exponential utility marginalized over variance $s^2$ distributed as Wishart (Gamma) distribution  $s^2\sim \Gamma(\frac{\alpha}{2},\frac{\sigma^2}{\alpha})$, we have the following equation:
\be 
x_c = \mu_c-\frac{a}{2} \sigma_0^2 +\frac{\alpha}{2 a} \ln \left[1-\frac{a^2}{\alpha}  \sigma_c^2\right] 
\ee
The CE values are typically determined through subjective evaluations of a gamble or investment. For example, consider an investment portfolio starting with an initial value of \$1 and two possible monetary outcomes: a payoff of \$1.21 (21\% gain) with probability $p_1=\frac{2}{3}$ and a payoff of \$0.9 (10\% loss) with probability $p_2=\frac{1}{3}$, and a given CE $x_c = \$1.07$ (7\% return). The expected value of this gamble is as follows:
\be
\mu_c = \frac{2}{3} \times 1.21 + \frac{1}{3} \times 0.9= 1.11,\,\,\, \sigma_c=\sqrt{\frac{2}{3}\times (1.21 - 1.11)^2 + \frac{1}{3} \times (0.9 - 1.11)^2} = 0.15
\ee
Using Eq.~\ref{eq:risk_a} with $\sigma_0 = 0$, we obtain a risk-aversion coefficient of $a = 3.4$. 

For typical S\&P 500 parameters ($\mu = 0.08$, $\sigma_p = 0.15$, and $r_0 = 0.02$) and $a = 3.4$, the exponential utility risk asset allocation is $w = \frac{1}{a}\frac{\mu - r_0}{\sigma_p^2} = 0.78$ and the allocation to the risk-free asset is $w_{r0} = 1 - w = 0.22$.

Once the risk-aversion coefficient $a$ is fixed, no remaining free parameters remain in the model, and its output becomes fully determined—enabling consistent application to decision-making scenarios. Most importantly, in contrast with the conventional approach,  the sum of weights need not to be normalized to one, and the solution provides allocations consistent with the chosen CE. 

\section{Optimal Leverage}
The Kelly criterion is a framework for optimizing capital growth by maximizing the expected logarithmic utility of wealth, \( E[\log X_T] \), over a sequence of $T$ gambles or investment decisions. Proposed by Kelly \cite{Kelly:1956}, the criterion provides an optimal fraction, \(f \), to allocate in each gamble, balancing growth with risk to avoid total financial ruin.

\subsection{Kelly criterion with logarithmic utility function}
Sequential gambling is a multiplicative process where the distribution of outcomes (wealth) naturally converges to a lognormal distribution, and GBM is an attractor process for a large class of multiplicative dynamics. A notable feature of GBM is that the mode of the wealth distribution is exponentially suppressed relative to the mean. The contribution to the mean is dominated by a few big winners, which are highly improbable but disproportionately large outcomes. Almost all trajectories of wealth eventually converge to zero (ruin) as time progresses if the process is volatile or has non-negligible drift uncertainty, despite an increase in average wealth. The proper choice of leverage helps mitigate this effect.
In this section, we investigate the optimal leverage problem in continuous and binary settings. In addition, we study the effect of uncertainty in expected return $\mu$ and variance $\sigma^2$. 

Suppose the dynamics of wealth $X_t$ is modeled by the following GBM: 
\be
\frac{d X_t}{X_t} = \mu \, dt + \sigma \, dW.
\label{eq:gbm_eq}
\ee
Solving Eq.~\ref{eq:gbm_eq}, the final wealth $X_T$ is given by the lognormal distribution $P(X_T) \sim LN(\mu_{\ln}, \sigma_{\ln}; X_T)$ where the logarithm of relative wealth is distributed as follows:
\be
\ln\left(\frac{X_T}{X_0}\right) \sim \mathcal{N}(\mu_{\ln}, \sigma^2_{\ln}), \,\,\,
\mu_{\ln} = \left(\mu - \frac{1}{2} \sigma^2\right) T, \,\,\, \sigma^2_{\ln} = \sigma^2 T
\label{eq:gbm_eq_ln}
\ee 
Assuming leverage $f$, an allocation between cash at rate $r_0$ and a risky portfolio with drift $\mu_r$ and volatility $\sigma_r$, we have the following:
\be
\mu = (1 - f) r_0 + f \mu_r=r_0 + f (\mu_r-r_0), \quad \sigma = f \sigma_r
\label{eq:musigma1}
\ee
We assume that optimal diversification is obtained in single-period optimization under exponential utility, which leads to the expected portfolio return $E[r_{\text{CARA}}]=r_0+\mu_p$ and variance $\operatorname{Var}[r_{\text{CARA}}]=\sigma_p^2$, obtained from Eq.~\ref{eq:allocation2}. The first two moments of the one-step GBM evolution in Eq.~\ref{eq:gbm_eq} with $\Delta t=1$ for the arithmetic return $r_{\text{GBM}}$ are given by $E[r_{\text{GBM}}]=\mu$ and $\operatorname{Var}[r_{\text{GBM}}]=\sigma^2$. We match both moments under the condition $f=1$, which gives $\mu_r - r_0 \approx \mu_p$ and $\sigma_r^2 \approx \sigma_p^2$.

The wealth $X_T$ is lognormally distributed. The average wealth grows as
\be
\text{mean } X_T = e^{\mu_{\ln} + \frac{1}{2} \sigma_{\ln}^2} = e^{\mu T}.
\ee
The typical wealth, represented by the mode, is given by
\be
\text{mode } X_T = e^{\mu_{\ln} - \sigma_{\ln}^2} = e^{\mu T - \frac{3}{2}\sigma^2 T}.
\ee
Note that the mode is exponentially suppressed by a factor of $e^{-\frac{3}{2} \sigma^2 T}$ relative to the mean. Also, in light of Eq.~\ref{eq:kelly_mu}, the mode goes to zero exponentially fast as $T$ increases and $\sigma_0 > 0$. 

In the GMV approach, we optimize the trade-off between the expected logarithmic utility and its variance:
\be
GMV= E\left[\ln \frac{X_T}{X_0} \right] - \frac{\lambda}{2} \operatorname{Var}\left[\ln \frac{X_T}{X_0}\right]=  \left(\mu - \frac{1}{2} \sigma^2\right)T - \frac{\lambda}{2} \sigma^2 T=f (\mu_r - r_0) T + r_0 T - \frac{1+\lambda}{2} f^2  \sigma_r^2 T 
\label{eq:gmb_kelly}
\ee
%
The optimal leverage is given by
\be
f^* =\arg \max_f GMV  =\frac{1}{1+\lambda} \frac{\mu_r-r_0}{\sigma_r^2}\approx \frac{1}{1+\lambda} \frac{\mu_p}{\sigma_p^2}
\label{eq:kelly1}
\ee
Assuming $\lambda = 1$, we obtain the \mbox{\it{half}}-Kelly criterion commonly used by practitioners:
\be
f^* = \frac{\mu_r-r_0}{2 \sigma_r^2} \approx  \frac{1}{2} \frac{\mu_p}{\sigma_p^2}
\label{eq:kelly_half}
\ee
while the original Kelly formula $f_K^* = \frac{\mu_r-r_0}{\sigma_r^2}$ can be derived from taking $\lambda = 0$ in Eq.~\ref{eq:kelly1}. Using Eq.~\ref{eq:kelly_half}, the half-Kelly leverage for typical S\&P 500 parameters, given excess return $\mu_p=0.06$ and volatility $\sigma_p=0.15$, is $f^*=1.33$.

As the next step, consider a model where $\mu$ is a normally distributed random variable, $\mu \sim \mathcal{N}(\mu_0, \sigma^2_0)$.
The wealth $X_T$ is distributed as follows:
\be
\tilde{P}(X_T) = \int_{-\infty}^{-\infty} d\mu \, \mathcal{N}(\mu_0, \sigma^2_0; \mu) \, P(\mu, \sigma; X_T).
\ee
where $P(X_T) = LN(\mu_{\ln}, \sigma_{\ln}; X_T)$. The integral can be calculated analytically and the logarithm of the wealth $X_T$ is again normally distributed:
\be
\ln\left(\frac{X_T}{X_0}\right) \sim \mathcal{N}(\tilde{\mu}_{\ln}, \tilde{\sigma}^2_{\ln}; X_T), \,\,\,\,
\tilde{\mu}_{\ln} = \left(\mu_0 - \frac{1}{2} \sigma^2\right) T,\,\,\, \quad \tilde{\sigma}^2_{\ln} = \sigma^2 T + \sigma_0^2 T^2.
\label{eq:kelly_mu}
\ee  
Note that the drift uncertainty $\sigma^2_0$ leads to a quadratic increase in the variance of the logarithmic utility. 
We use  decomposition 
$\mu_0 = (1 - f) r_0 + f \mu_{0r}$ and  $\sigma = f \sigma_r$,  with estimates from Eq.~\ref{eq:allocation2}: $\mu_{0r}=\mu_p $, $\sigma^2_0=\sigma^2_{0p}$ and $\sigma^2=\sigma^2_p$. The GMV criterion is given by
\be
GMV= E\left[\ln \frac{X_T}{X_0} \right] - \frac{\lambda}{2}  \operatorname{Var}\left[\ln \frac{X_T}{X_0}\right]=f (\mu_{0r} - r_0) T + r_0 T -  \frac{(1 + \lambda)}{2} f^2 (\sigma_r^2+\sigma_0^2 T) T
\label{eq:kelly_gmv}
\ee 
The optimal leverage is given by
\be
f^* =\arg \max_f GMV  =\frac{1}{1+\lambda} \frac{\mu_{0r}-r_0}{( \sigma_r^2+\sigma_0^2 T)}.
\label{eq:kelly2}
\ee
Uncertainty in the drift causes optimal leverage to depend on the time horizon. The longer the gambler needs to survive, the smaller the optimal leverage should be. $\sigma_0^2 T$ term in Eq.~\ref{eq:kelly2} challenges the theorems about the optimality of the Kelly betting as they often assume $T \to \infty$.

To understand the effect of the uncertain variance of the portfolio $\sigma^2$, we  model it as a gamma distributed random variable, $s^2\sim \Gamma(\frac{\alpha}{2}, \frac{2 \sigma^2}{\alpha})$. If $X_T$ is distributed as $P(X_T)= LN(\bar \mu_{\ln}, \bar \sigma_{\ln}; X_T)$  with $\bar \mu_{ln}=(\mu-\frac{1}{2} s^2) T$ and $\bar\sigma^2_{ln} =  s^2 T$, the distribution of wealth with uncertain variance $\tilde{P}(X_T)$ is given by
\be
\tilde{P}(X_T) = \int_0^{\infty} d s^2 \, \Gamma(\frac{\alpha}{2}, \frac{2 \sigma^2}{\alpha};s^2) \, LN(\bar \mu_{ln}, \bar \sigma_{ln}; X_T)=
\label{eq:gmb_sigma_a}   
\ee
$$
\frac{\alpha ^{\alpha /2} e^{\mu T/2} \left(4 \alpha +\sigma ^2 T\right)^{\frac{1-\alpha}{4}}  | T \mu -\log (X_T)| ^{\frac{\alpha -1}{2}} K_{\frac{\alpha -1}{2}}\left(\frac{\sqrt{4 \alpha+\sigma ^2 T } | T \mu -\log (X_T)| }{2 \sqrt{T} \sigma }\right)}{\sqrt{\pi }   \Gamma
   \left(\frac{\alpha }{2}\right) (\sigma  \sqrt{T})^{\frac{1+\alpha }{2}}  X_T^{\frac{3}{2}}}
$$
Taking the final wealth distribution $\tilde P(X_T)$ from Eq.~\ref{eq:gmb_sigma_a} and the definition of $\mu$ and $\sigma$ from Eq.~\ref{eq:musigma1}, we calculate numerically the mean $\mu_{\sigma}=\int_0^\infty d X_T \tilde P(X_T) \ln \frac{X_T}{X_0}$ and variance \\
$\sigma^2_{\sigma}=\int_0^\infty d X_T  \tilde P(X_T) (\ln \frac{X_T}{X_0}-\mu_{\sigma})^2$ of the logarithmic utility and find the optimal leverage $f^*$ as the solution of the GMV problem 
$f^*=\arg\max_f [\mu_{\sigma}-\frac{\lambda}{2} \sigma^2_{\sigma}]$. 
%

An alternative to the continuous GBM approximation of compounding in Eq.~\ref{eq:gbm_eq} is to directly represent the multiplicative betting process as a discrete process: $X_t = X_{t-1}\,\exp\left(\mu_d + \sigma_d\,\varepsilon_t\right),\, \varepsilon_t \sim N(0,1)$.
The wealth \(X_T\) is lognormally distributed $\ln X_T/X_0 = \sum_{i=1}^T \left(\mu_d + \sigma_d\,\varepsilon_i\right)=N\left(\mu_d T,\sigma_d^2 T\right).$ The parameters $\mu_d$ and $\sigma_d$ can be calibrated on empirical arithmetic return $r_A$ using the following equations:
$$
    E[r_A] = E[\frac{X_t}{X_{t-1}} - 1] = 
    \exp\left(\mu_d + \frac{\sigma_d^2}{2}\right) - 1, \,\, 
\operatorname{Var}[r_A] = \exp\left(2\mu_d + \sigma_d^2\right)\left(\exp\left(\sigma_d^2\right) - 1\right).
$$
For small return  levels, we have  $\sigma^2_d \approx \operatorname{Var}[r_A])$ and $\mu_d \approx \ln\left(1 + E[r_A]\right) - \frac{\operatorname{Var}[r_A]}{2}$. Thus, the discrete model essentially reproduces the solution of the GBM equation.
 
Results from Sections 3 and 4 can be used independently or combined. In the latter case, we combine the optimal weights $\bs w^*$ obtained from the optimization of exponential utility with the optimal leverage $f^*$ obtained from the optimization of the logarithmic utility of wealth, thereby yielding the final allocation weights $\bs w_f^* = f^* \bs w^*$. This approach to allocation is conceptually similar to the model predictive control (MPC) framework, where the cost functions are given by exponential utility for cross-asset allocation (optimal diversification) and the logarithmic utility of wealth for time diversification (optimal leverage). The control inputs are the allocation weights, and the system dynamics are explicitly defined by the portfolio-return model.


The derivation of GMV for binary betting, in both the deterministic and probabilistic cases, is presented in Appendix F. The Kelly criterion with a power utility function is discussed in Appendix G. We discuss the practical considerations in Appendix E.
   
\section{Conclusions}

In this paper, we study the problem of optimal diversification and leverage in a utility-based approach. 
We show that exponential utility generalizes the MV approach, offering a diversification rule while maintaining analytical tractability for cases such as ALD-distributed returns and Gaussian returns with an uncertain covariance matrix. An allocation derived from the exponential utility is calibrated based on the CE of the trade and has no free parameters left. 
The optimal leverage can be obtained from the logarithmic utility in the GMV approach, which balances the expected utility and its variance. This  naturally explains the half-Kelly criterion. Thus, we derive both relative allocation (diversification) and absolute allocation (leverage) within a unified framework.

We enhance existing methods by introducing a compound distribution of the outcome and the GMV. The use of the compound distribution and marginalization over parameters formally coincides with the posterior predictive distribution of the outcome, allowing us to use Bayesian machinery when modeling the outcome of an investment. The GMV can be seen as a generalization of the maximum expected utility principle and enables control over the dispersion of utility—which is particularly relevant in situations where the probability of ruin is non-negligible.

The proposed framework allows for a natural treatment of uncertainty in both the expected return and the covariance matrix, as well as arbitrary distributions of returns, although analytical results can only be obtained for a limited number of distributions. In summary, the framework links the utility formulation, parameter estimation using Bayesian methods, and, finally, the convex optimization of the resultant objective function.


\section{Appendix A: GMV examples}

In the standard economic approach, asset allocations are derived by maximizing the expected value of the investor's utility function: 
\be
w = \arg\max_w  E[U(X|D)] 
\label{eq:mv1a}
\ee
The utility $U(X)$ non-linearly transforms the investor's wealth $X$ and encodes his aversion to risk. The most commonly used utility functions are linear, logarithmic, exponential, and power utilities.

Although the MEU is a widely used solution for the rational behavior of economic agents, it has limitations that often manifest as paradoxes, such as the Ellsberg and Allais paradoxes. The MEU ignores the variability in utility, focusing only on average utility without considering fluctuations across scenarios. In this paper, we investigate the GMV framework:  
\be
w = \arg\max_w \left( E[U(X|D)] - \frac{\lambda}{2} \, \text{Var}[U(X|D)] \right)
\label{eq:gmv1a}
\ee
which addresses this by incorporating utility dispersion, capturing investors' aversion to utility instability.  Risk-averse investors may prefer portfolios with more stable utility, even if expected utility is slightly lower. 
Eq.~\ref{eq:gmv1a} can be seen as a generalization of the MEU approach.

We find that the variance term particularly relevant when using logarithmic utility applied to proportional wealth changes with sequential betting, where the probability of ruin is involved, as it explains the commonly used half-Kelly criterion for optimal leverage. 

We combine the GMV with a compound distribution of parameters, which naturally allows us to incorporate uncertainties.
Our view is that the approach provides a flexible framework to incorporate arbitrary outcome distributions and uncertainties into the decision-making process.

In this section, we discuss three examples of the GMV to better understand the role of the variance of utility $\text{Var}[U(X|D)]$. 

\subsection*{Example 1: Taylor expansion of a weakly non-linear utility}
In the univariate case with known parameters $\mu$ and $\sigma$, the expected value and variance of a utility can be derived using the Taylor expansion of a random variable. 
For an arbitrary utility function $U(X)$ of a random variable $X$, we can approximate the expectation and variance as follows:
\be 
E[U(X)] \approx U(\mu_X) + \frac{U''(\mu_X)}{2} \sigma_X^2,\,\, \operatorname{Var}\left[U(X)\right] \approx \left(U'(\mu_{X})\right)^{2} \sigma_{X}^{2} + \frac{1}{2} \left(U''(\mu_{X})\right)^{2} \sigma_{X}^{4}
\label{eq:eu_appr}
\ee
where $\mu_X = E[X]$ and $\sigma_X^2 = \text{Var}[X]$.                         

The expected value has a term proportional to the second derivative $U''(\mu_X)$, which reflects risk aversion. The variance contribution includes a term proportional to the square of the first derivative $(U'(\mu_X))^2$, corresponding to marginal utility. It is clear that the variance contribution has a positive effect on risk.

We derive the GMV for multivariate Gaussian returns following \cite{Rego:2021}. Let $\boldsymbol{X} \sim \mathcal{N} 
(\boldsymbol{\mu}, \boldsymbol{\Sigma})$ and $\boldsymbol{\delta} = \boldsymbol{X} - \boldsymbol{\mu}$. Further, let $U(\boldsymbol{X})$ be a nonlinear scalar function of    $\boldsymbol{X}$, with gradient vector $\boldsymbol{g}$ and Hessian matrix $\boldsymbol{H}$ at $\boldsymbol{X} = \boldsymbol{\mu}$, such that 
\be
g_i = \frac{\partial U}{\partial x_i} \bigg|_{x_i = \mu_i} \quad  
\text{and} \quad H_{ij} = \frac{\partial^2 U}{\partial x_i \partial x_j}\bigg|_{[x_i, x_j] = [\mu_i, \mu_j]}.
\label{eq:gh}
\ee
The second-order Taylor series expansion of $U(\boldsymbol{X})$ centered at $\boldsymbol{\mu}$ is then
\be
U(\boldsymbol{X}) \approx U(\boldsymbol{\mu}) + \boldsymbol{g}^T \boldsymbol{\delta} + \frac{1}{2} \boldsymbol{\delta}^T \boldsymbol{H} \boldsymbol{\delta}.
\ee
From Eq.~\ref{eq:gh}, we can explicitly approximate the expected value and variance as follows:
\be
E[U(\boldsymbol{X})] \approx U(\boldsymbol{\mu}) + \frac{1}{2} \mathrm{tr}(\boldsymbol{H} \boldsymbol{\Sigma}),\,\,
\mathrm{Var}[U(\boldsymbol{x})] \approx \boldsymbol{g}^T \boldsymbol{\Sigma} \boldsymbol{g} + \frac{1}{2} \mathrm{tr}((\boldsymbol{H} \boldsymbol{\Sigma})^2)
\ee
Assuming a weakly non-linear utility, $U(\boldsymbol{\mu}) \approx \boldsymbol{\mu}$, we obtain a convex approximation to the GMV.
\subsection*{Example 2: Linear utility}
Consider the simplest example of applying a compounding distribution $H$ with an uncertain expected return $\mu$ to a linear utility function $U(X)=X$:
\be 
X \sim P(\mu, \sigma), \,\,\, \mu \sim G_{\mu}(\mu_0, \sigma_0)
\ee
The first two moments of the compound distribution $H$ are given by the law of total expectation and the law of total variance:
\be 
E_H[X] = E_{G_{\mu}} \left[ E_P[X|\mu] \right]=\mu_0
\ee
$$
\text{Var}_H(X) = E_{G_{\mu}} \left[ \text{Var}_P(X|\mu) \right] + \text{Var}_{G_{\mu}} \left[ E_P[X|\mu] \right]=\sigma^2 + \sigma_0^2
$$
The generalized mean variance objective of the linear utility is given by
\be
GMV_{lin}=E_H[U[X]]-\frac{\lambda}{2} \text{Var}_H[U[X]]= \mu_0-\frac{\lambda}{2}(\sigma^2 + \sigma_0^2)
\ee
Correspondingly, the optimal weight $w^*_{lin}$ for a risk asset MV allocation is given by
\be
w^*_{lin}=\frac{1}{\lambda}\frac{\mu_0}{\sigma^2+\sigma_0^2}
\ee

\subsection*{Example 3: Interpretation of the variance term}
Consider a random variable $X \sim N(\mu, \sigma)$ with an uncertain mean $\mu \sim G_{\mu}(\mu_0,\sigma_0 \mid D)$. The expected utility is given by
\be
E[U(X|D)] = E_{P}[E_{G_{\mu}}[U(X|\mu)]] =  \int dX \, U(X) \int d\mu \,  P(X|\mu) G_{\mu}(\mu|D) 
\ee
By the law of total variance, the variance of $U(X|D)$ given data D is given by
\be 
\text{Var}[U(X|D)]= E_{G_{\mu}}[\text{Var}_P[U(X|\theta)]] + \text{Var}_{G_{\mu}}[E_P[U(X|\theta)]]
\ee
The first term can be interpreted as the average variance, taking into account the distribution of parameters $G_{\mu}$. The second term represents the variability contributed by the random variable $G_{\mu}$ to the location of expectation of the utility.
Another interpretation of this decomposition arises when the expected return is considered a random variable with two-states (or regimes). In this case, the first term represents the average variance within each individual regime (within group variance), while the second term represents the variance across different regimes (between group variance). 

\section{Appendix B: Analytical results for mean and variance of utility functions}
In this section, we calculate the mean and variance of the most commonly used utility functions, which are used in the GMV objective:
\be
w^*=\arg\max_w \left(E[U(X)]-\frac{\lambda}{2} \text{Var}[U(X)]\right)
\ee

Throughout the paper, we use the equivalent notation for the normal distribution: $N(\mu, \sigma) \equiv \mathcal{N}(\mu, \sigma^2)$. Also, we interchangeably use the terms return and investment outcome when they are trivially connected.

\subsection*{Linear utility}
The linear utility is defined as follows:
\be 
U_{\text{lin}}(x)=x
\ee
Assuming returns are Gaussian $X\sim N(\mu,\sigma)$, we have: 
\be
E[U_{\text{lin}}(X)]=  
 \int_{-\infty}^{\infty} N(\mu, \sigma; x) U_{\text{lin}}(x) \, dx =\mu 
\ee
\be
E[U_{\text{lin}}(X)^2] =   
\int_{-\infty}^{\infty} N(\mu, \sigma; x) U^2_{\text{lin}}(x) \, dx =\mu^2 + \sigma^2
\ee
\be
\text{Var}[U_{\text{lin}}(X)] =E[U_{\text{lin}}(X)^2]-E[U_{\text{lin}}(X)]^2=\sigma^2
\ee
Assuming returns are Gaussian $X\sim N(\mu,s)$ with $\mu\sim N(\mu_0, \sigma_0)$ and $s^2\sim \Gamma\left( \frac{\alpha}{2}, \frac{2 \sigma^2}{\alpha}\right)$, we have: 
\be
E[U_{\text{lin}}(X)]= \int_0^{\infty}  \, ds^2 \Gamma\left( \frac{\alpha}{2}, \frac{2 \sigma^2}{\alpha}; s^2 \right)  \int_{-\infty}^{\infty}  d\mu \,
N(\mu_0, \sigma_0; \mu) 
 \int_{-\infty}^{\infty} N(\mu, \sqrt{s^2}; x) \cdot x \, dx =\mu_0 
\ee
\be
\text{Var}[U_{\text{lin}}(X)] =\sigma^2 + \sigma_0^2
\ee
The GMV objective is given by
\be
GMV=\mu_0 +\frac{\lambda}{2}( \sigma^2 + \sigma_0^2)
\ee
\subsection*{Exponential utility: Univariate case}
The exponential utility (CARA) is given by
\be
U_{\text{a}}(x,a) = \frac{1 - \exp(-a x)}{a}, \quad \text{if } a \neq 0
\ee
Assuming returns are Gaussian $X\sim N(\mu,\sigma)$ and no uncertainty in $\mu$ and $\sigma$, we have:
\be
E[U_a(X)] = \int_{-\infty}^{\infty} N(\mu, \sigma; x)  U_{a}(w x, a) \, dx 
=\frac{1 - e^{\frac{1}{2} a w (-2 \mu + a w \sigma^2)}}{a}
\ee
\be
\text{Var}[U_a(X)] = \frac{e^{a w (-2 \mu + a w \sigma^2)} \left(-1 + e^{a^2 w^2 \sigma^2}\right)}{a^2}
\ee
where we keep the risk-asset allocation weight $w$ explicit.
In case $\mu\sim N(\mu_0, \sigma_0)$, we have :
\be
E[U_a(X)] = \int_{-\infty}^{\infty} d\mu \, N(\mu_0, \sigma_0; \mu) \int_{-\infty}^{\infty}  dx \, N(\mu, \sigma; x)  U_{a}(w x, a) \,=
\ee
$$
=\frac{1 - e^{\frac{1}{2} a w \left(-2 \mu_0 + a w (\sigma^2 + \sigma_0^2)\right)}}{a}
$$
\be
\text{Var}[U_a(X)] =\frac{-e^{a w \left(-2 \mu_0 + a w (\sigma^2 + \sigma_0^2)\right)} + e^{2 a w \left(-\mu_0 + a w (\sigma^2 + \sigma_0^2)\right)}}{a^2}
\ee
Assuming returns are Gaussian $X\sim N(\mu,s)$ with $\mu\sim N(\mu_0, \sigma_0)$ and $s^2\sim \Gamma\left( \frac{\alpha}{2}, \frac{2 \sigma^2}{\alpha}\right)$, we have: 
\be
E[U_a(X)]= \int_0^{\infty}  ds^2 \, \Gamma\left( \frac{\alpha}{2}, \frac{2 \sigma^2}{\alpha}; s^2 \right)   \int_{-\infty}^{\infty}  d\mu \, 
N(\mu_0, \sigma_0; \mu) 
\int_{-\infty}^{\infty} N(\mu, \sqrt{s^2}; x)  U_{a}(w x, a) \, dx =\, 
\ee
$$
=\frac{1 - e^{\frac{1}{2} a w \left(-2 \mu_0 + a w \sigma_0^2\right)} \left(\frac{\alpha}{\alpha - a^2 w^2 \sigma^2}\right)^{\frac{\alpha}{2}}}{a}
$$
\be 
\text{Var}[U_a(X)] =\frac{e^{a w \left(-2 \mu_0 + a w \sigma_0^2\right)} \left( e^{a^2 w^2 \sigma_0^2} \left(\frac{\alpha}{\alpha - 4 a^2 w^2 \sigma^2}\right)^{\frac{\alpha}{2}} - \left(\frac{\alpha}{\alpha - a^2 w^2 \sigma^2}\right)^{\alpha} \right)}{a^2}
\ee

\subsection*{Exponential utility: Multivariate case}

Suppose an investor's opportunity set consists of $N$ risky assets and a single risk-free asset. The risk-free asset's return for the period is $r_0 \geq 0$. We assume that the vector of returns $\bs r = (r_1, \ldots, r_N)^T$ of the risky assets follows either a multivariate normal distribution, $\bs r \sim N(\bs \mu, \bs \Sigma)$, or an ALD distribution, $\bs r \sim \text{ALD}(\bs \mu, \bs \Sigma, \bs \mu_a)$. Given initial wealth $W_0 > 0$, the investor chooses portfolio weights $\bs w$ for the risky assets such that the expected utility $E[U(W)]$ of next period's wealth is maximized. The next period's wealth is given by
\be 
W = W_0 \left(1 + (1 - \bs w^T \boldsymbol{I})r_0 + \bs w^T \bs r \right),
\ee
and the optimization problem is given by
\be 
\bs w^*=\arg \max_{\bs w} E\left[U\left(W_0 \left(1 + (1 - \bs w^T \boldsymbol{I})r_0 + \bs w^T \bs r \right)\right)\right].
\ee
It can be rewritten as follows:
\be
\bs w^*=\arg \max_{\bs w} e^{-a W_0 (1 + (1 - \bs w^T \boldsymbol{I})r_0)} E \left[-e^{-a W_0 \bs w^T \bs r}\right]
\ee
For our purposes, it is sufficient to consider $W_0 = 1$. Thus, the nontrivial part of the optimization of exponential utility is given by the term $E\left[-e^{-a \bs w^T \bs r}\right]$.
The expected utility function $E_{ALD}[U_a]$ for ALD distributed return with uncertain location parameter $\bs \mu$ is given by \cite{MarkovA:2023}
\be
E_{ALD}[-e^{-a \bs w^T \bs r}] = (-1) e^{-a \boldsymbol \mu_0^T \boldsymbol w+\frac{a^2}{2}  \boldsymbol w^T \boldsymbol \Sigma_0 \boldsymbol w-\ln \left[1-\frac{a^2 \boldsymbol w^T \boldsymbol \Sigma \boldsymbol w }{2}+a  \boldsymbol \mu_a^T \boldsymbol w\right]}
\label{ald-weight-marg}
\ee
The expected utility function $E_{\boldsymbol \mu,\boldsymbol \Sigma}[U_a]$ for multivariate Gaussian returns with uncertain $\bs \mu$ and $\bs \Sigma$, after marginalization over the covariance matrix, is given by \cite{MarkovB:2023}
\be 
E_{\boldsymbol \mu,\boldsymbol \Sigma}[-e^{-a \bs w^T \bs r}]  
= (-1)e^{-a \boldsymbol  \mu_0^T \boldsymbol  w+\frac{a^2}{2}  \boldsymbol w^T \boldsymbol  \Sigma_0 \boldsymbol w-\frac{\alpha}{2} \ln \left[1-\frac{a^2}{\alpha}  (\boldsymbol w^T \boldsymbol \Sigma \boldsymbol w)\right]}
\label{normal-eu-1a} 
\ee
The results above are sufficient to calculate the GMV objective. Denote $f(a)= E[-e^{-a \bs w^T \bs x}]$, and we have:
\be
U_{a}(x,a)=\frac{1-e^{-a wx}}{a},\,\,\, \,\, U_{a}^2(x,a)=\frac{1}{a^2}-\frac{2 e^{-a w x}}{a^2}+\frac{e^{-2 a w x}}{a^2}
\ee
Thus, it is enough to know $f(a)$ to calculate $E[U_{a}]=\frac{1+f(a)}{a}$ and  $E[U_{a}^2]=\frac{1}{a^2}+\frac{2 f(a)}{a^2}-\frac{f(2 a)}{a^2}$ and variance $\text{Var}[U_{a}]=E[U_{a}^2]-E[U_{a}]^2$. 
The practical application of the GMV of exponential utility for portfolio allocation is constrained by the fact that the variance term violates the convexity of the objective.

\subsection*{Logarithmic utility}
The logarithmic utility is defined as follows:
\be 
U_{\text{log}}(x)=\ln x
\ee
Assuming the outcome is lognormally distributed $X\sim LN(\mu,\sigma)$ with $\mu\sim N(\mu_0, \sigma_0)$, we have:
\be
E[U_{\text{log}}(X)]= \int_{-\infty}^{\infty} d\mu\,\, N(\mu_0, \sigma_0; \mu) \int_0^{\infty} LN(\mu, \sigma; x)  U_{\text{log}}(x) \, dx \, =\mu_0
\ee
%
\be
\text{Var}[U_{\text{log}}(X)]  = \sigma^2 + \sigma_0^2
\ee
Assuming the outcome is lognormally distributed $X\sim LN(\mu,s)$ with $\mu\sim N(\mu_0, \sigma_0)$
and $s^2\sim \Gamma\left( \frac{\alpha}{2}, \frac{2 \sigma^2}{\alpha}\right)$, we have :
\be
E[U_{\text{log}}(X)] = \int_0^{\infty}  ds^2 \,  \Gamma\left( \frac{\alpha}{2}, \frac{2 \sigma^2}{\alpha}; s^2 \right)  \int_{-\infty}^{\infty}  d\mu \,
N(\mu_0, \sigma_0; \mu) 
\int_0^{\infty} dx \, LN(\mu, \sqrt{s^2}; x) \cdot  U_{\text{log}}(x) \,=\mu_0
\ee
%
\be
Var[U_{\text{log}}(X)]  = \sigma^2 + \sigma_0^2
\ee

\subsection{Power utility}
The power utility (CRRA) is defined as follows:
\be 
U_r(x,\gamma) = 
        \frac{x^{1-\gamma}-1}{1-\gamma}, \,\, \gamma>0   ,\,\, \gamma \neq 1
\ee  
Assuming the outcome is lognormally distributed $X\sim LN(\mu,\sigma)$,   
the expected utility and its variance are given by:
\be
E[U_r(X)]= \int_0^{\infty} LN(\mu, \sigma; x)   U_{r}(x, \gamma) \, dx=\frac{ e^{\frac{1}{2} (1 - \gamma) \left(2 \mu + (1 - \gamma) \sigma^2\right)}-1}{1 - \gamma}
\ee
\be
\text{Var}[U_r(X)]= \frac{e^{2 \mu - 2 \gamma (\mu + 2 \sigma^2)} \left(e^{2 (1 + \gamma^2) \sigma^2}-e^{(1 + \gamma)^2 \sigma^2}  \right)}{(1 - \gamma)^2}
\ee
In case $\mu\sim N(\mu_0, \sigma_0)$, we have: 
\be
E[U_r(X)] =\int_{-\infty}^{\infty}d\mu N(\mu_0, \sigma_0; \mu) \int_0^{\infty} LN(\mu, \sigma; x)  U_{r}(x, \gamma) \, dx \, =
\ee
$$
=\frac{ e^{\frac{1}{2} (1 - \gamma) \left(2 \mu_0 + (1 - \gamma) (\sigma^2 + \sigma_0^2)\right)}-1}{1 - \gamma}
$$
%
\be
\begin{aligned}
\text{Var}[U_r(X)] 
&= \frac{
    e^{(1 - \gamma)\bigl[\,2\mu_0 
       + (1 - \gamma)\,(\sigma^2 + \sigma_0^2)\bigr]}
   }
   {2\,(1 - \gamma)^2}
\\[6pt]
&\quad\times 
\Bigl(
  -2 
  + e^{(1 - \gamma)^2(\sigma^2 + \sigma_0^2)}
    \Bigl(
      1 
      + \mathrm{Erfc}\!\Bigl(\tfrac{\mu_0 
               + 2(1 - \gamma)\,\sigma_0^2}{\sqrt{2}\,\sigma_0}\Bigr)
      + Q\!\Bigl(
          \tfrac12,\,0,\,
          \tfrac12\bigl(2 - 2\gamma 
                        + \tfrac{\mu_0}{\sigma_0^2}\bigr)^2
                        \sigma_0^2
        \Bigr)
    \Bigr)
\Bigr).
\end{aligned}
\ee
where $\operatorname{Q}(a,z_0,z_1)$ is the regularized gamma function and $\operatorname{Erf}(x)$ is the error function.

\section{Appendix C: Analytical solutions for exponential utility-based allocations}
\subsection*{Analytical solution for ALD distributed return}
For multivariate ALD returns $\bs r \sim ALD(\bs \mu,\bs \Sigma,\bs \mu_a)$ with uncertain location vector $\boldsymbol\mu\sim \mathcal{N} (\boldsymbol \mu_0,\boldsymbol \Sigma_0)$, we have the following optimization objective \cite{MarkovA:2023}:
\be
\boldsymbol w^*=\arg \max_{\boldsymbol w} \left[  (1-\boldsymbol w^T \boldsymbol I) r_0+ \boldsymbol \mu_0^T \boldsymbol w-\frac{a}{2}  \boldsymbol w^T \boldsymbol \Sigma_0 \boldsymbol w+\frac{1}{a} \ln\left[1-\frac{a^2}{2} \boldsymbol  w^T \boldsymbol \Sigma \boldsymbol w+a  \boldsymbol \mu_a^T \boldsymbol w\right] \right]
\label{eq:opt-weight4a}
\ee
such that $\frac{a^2}{2} \boldsymbol{w}^T \boldsymbol{\Sigma} \boldsymbol{w} - a \boldsymbol{\mu}_a^T \boldsymbol{w} < 1$.
Here $\bs \Sigma_0$ is a diagonal matrix of variances of the expected return $\sigma_0^2$, $\mathbf{\Sigma}$ is a scale matrix , and $\boldsymbol{\mu_a}$ is an asymmetry parameter.  

If the matrix $\boldsymbol{\Sigma}_0 = 0$ and there are no constraints, it is possible to solve the equation analytically. 
We have the following optimization problem:
\be
\boldsymbol{w}^* = \arg \max_{\boldsymbol{w}} \left[ (1 - \boldsymbol{w}^T \boldsymbol{I}) r_0 + \boldsymbol{\mu}^T \boldsymbol{w} + \frac{1}{a} \ln\left(1 - \frac{a^2}{2} \boldsymbol{w}^T \boldsymbol{\Sigma} \boldsymbol{w} + a \boldsymbol{\mu}_a^T \boldsymbol{w}\right) \right]
\label{eq:multi-ald-weightA}
\ee
Maximization of Eq.~\ref{eq:multi-ald-weightA} results in the following system of quadratic equations with respect to $\boldsymbol{w}$:
\be
(\boldsymbol{\mu} - r_0 \boldsymbol{I}) \left(1 - \frac{a^2}{2} \boldsymbol{w}^T \boldsymbol{\Sigma} \boldsymbol{w} + a \boldsymbol{\mu}_a^T \boldsymbol{w}\right) + (\boldsymbol{\mu}_a - a \boldsymbol{\Sigma} \boldsymbol{w}) = 0
\label{eq:ald-q2}
\ee
We employ the following ansatz for the allocation weights $\boldsymbol{w}$:
\be
\boldsymbol{w} = \frac{g_{ALD}}{a} \boldsymbol{\Sigma}^{-1} (\boldsymbol{\mu} - r_0 \boldsymbol{I}) + \frac{1}{a} \boldsymbol{\Sigma}^{-1} \boldsymbol{\mu}_a
\ee
Substituting the ansatz into Eq.~\ref{eq:ald-q2}, we have the following equation for $g_{ALD}$:
\be
\frac{q}{2}g_{ALD}^2  + g_{ALD} - \left(1 + \frac{1}{2} v\right) = 0
\ee
where $q = (\boldsymbol{\mu} - r_0 \boldsymbol{I})^T \boldsymbol{\Sigma}^{-1} (\boldsymbol{\mu} - r_0 \boldsymbol{I})$ and $v = \boldsymbol{\mu}_a^T \boldsymbol{\Sigma}^{-1} \boldsymbol{\mu}_a$. The solution is given by
\be
g_{ALD} = \frac{-1 + \sqrt{1 + 2 q + q v}}{q}
\ee
The optimal portfolio allocation for ALD returns consists of three components: a risk-free allocation to cash $(1 - \boldsymbol{w}^T \boldsymbol{I}) r_0$, and two risky portfolios, $\frac{g_{ALD}}{a}\boldsymbol{\Sigma}^{-1} (\boldsymbol{\mu} - r_0 \boldsymbol{I})$ and $\frac{1}{a}\boldsymbol{\Sigma}^{-1} \boldsymbol{\mu}_a$.

\subsection*{Analytical solution for Gaussian return with Wishart covariance noise}
For normally distributed returns $\bs r \sim N(\bs \mu,\bs S)$ with uncertain location vector $\boldsymbol\mu\sim \mathcal{N} (\boldsymbol \mu_0,\boldsymbol \Sigma_0)$ and uncertain covariance matrix $\bs S \sim W_N(\alpha,\frac{\bs \Sigma}{\alpha})$, the optimal weights $\boldsymbol  w^*$ after marginalization over $\bs \mu$ and $\bs S$  are given by \cite{MarkovB:2023}
\be
\boldsymbol w^*=\arg \max_{\boldsymbol w} \left[ (1-\boldsymbol w^T \boldsymbol I) r_0+ \boldsymbol \mu_0^T \boldsymbol w-\frac{a}{2}  \boldsymbol w^T \boldsymbol  \Sigma_0 \boldsymbol w +\frac{\alpha}{2 a} \ln \left[1-\frac{a^2}{\alpha}  (\boldsymbol w^T \boldsymbol  \Sigma \boldsymbol w)\right] \right]
\label{eq:opt-weight3aA}
\ee
Here $\alpha$ is a parameter that controls the noise level of the covariance matrix. 

Maximizing Eq.~\ref{eq:opt-weight3aA} in the case of zero uncertainty in the expected returns $ \boldsymbol{\Sigma}_0 = 0$, yields a system of quadratic equations with respect to $ \boldsymbol{w} $:
\be
(\boldsymbol{\mu_0} - r_0 \boldsymbol{I})  \left (1 -\frac{a^2}{\alpha} \boldsymbol  w^T \boldsymbol \Sigma \boldsymbol w \right)-a \boldsymbol \Sigma \boldsymbol w=0 
\label{eq:wishart-q2}
\ee
Substituting an ansatz 
\begin{equation}
\boldsymbol w^* =g_W\times \left(\frac{1}{a}\boldsymbol \Sigma^{-1} \boldsymbol (\boldsymbol{\mu_0} - r_0 \boldsymbol{I}) \right)
\label{eq:wishart-q3}
\end{equation}
to Eq.~\ref{eq:wishart-q2}, the scaling factor is given by
\be 
g_W= \frac{1}{2 q} \left(\sqrt{\alpha(4 q+\alpha)}-\alpha\right)
\ee
where $q = (\boldsymbol{\mu_0} - r_0 \boldsymbol{I})^T \boldsymbol{\Sigma}^{-1} (\boldsymbol{\mu_0} - r_0 \boldsymbol{I})$. 

\section{Appendix D: Brownian motion with uncertain parameters} 
In this section, we investigate ABM and GMB models with uncertain drift and variance. The main result is a scaling of the variance of returns, which arises from the non-stationarity of the drift and the uncertainty in estimating the initial drift.
\subsection*{Arithmetic Brownian motion with uncertain drift: continuous-time model}
Consider an arithmetic Brownian motion \(X_t\) defined by
\begin{equation}\label{eq:ABM_cont}
X_t = X_0 + \int_0^t \mu_s\,ds + \sigma\,W_t,
\end{equation}
where \(X_0\) is the initial value, \(W_t\) is a standard Brownian motion, and \(\sigma\) is a volatility parameter.
The drift process is assumed to be stochastic:
\begin{equation}\label{eq:drift_cont}
\mu_t = \mu_0 + \sigma_\mu\,B_t,
\end{equation}
where \(B_t\) is a Brownian motion independent of \(W_t\) and \(\sigma_\mu\) measures the volatility (uncertainty) in the drift. Moreover, the initial drift is assumed random $\mu_0 \sim N(\mu_{pd}, \sigma_{pd}^2)$.
Substituting Eq.~\ref{eq:drift_cont} into Eq.~\ref{eq:ABM_cont}, we obtain
\be
X_t = X_0 + \mu_0 t + \sigma_\mu \int_0^t B_s\,ds + \sigma\,W_t.
\ee
Because \(X_0\), \(\mu_0\), \(\int_0^t B_s\,ds\) and \(W_t\) are mutually independent, and all these components are Gaussian, we can compute the mean and variance.
The mean is given by
\[
E[X_t] = E[X_0] + t\,E[\mu_0] + \sigma_\mu\,E\left[\int_0^t B_s\,ds\right] + \sigma\,E[W_t] 
= X_0 + \mu_{pd}\,t.
\]
The variance of \(X_t\) is the sum of the independent contributions.The term \(\mu_0t\) has variance \(t^2 \sigma_{pd}^2\). The term \(\sigma\,W_t\) has variance \(\sigma^2 t\). The term \(\sigma_\mu I_t\) has variance \(\sigma_\mu^2\,t^3/3\). It follows from the fact that the stochastic integral   $I_t = \int_0^t B_s\,ds$ is Gaussian with zero mean and variance $\operatorname{Var}(I_t)=\frac{t^3}{3}.$
Thus,
\be 
\operatorname{Var}(X_t) = \sigma^2 t +\sigma_{pd}^2  t^2 +  \frac{\sigma_\mu^2 t^3}{3}.
\label{eq:eq_var}
\ee
The distribution of \(X_t\) is given by
\be 
X_t \sim \mathcal{N}\Bigl(X_0 + \mu_{pd}\,t,\;\sigma^2 t+ \sigma_{pd}^2 t^2   + \frac{\sigma_\mu^2 t^3}{3}\Bigr).
\ee
The increment is given by 
\be 
\Delta X_t=X_{t+\Delta}-X_t=\mu_0 \Delta+\sigma_\mu \int_t^{t+\Delta} B_s\,ds+\sigma\,(W_{t+\Delta}-W_{t})
\ee 
and is Gaussian with mean and variance
\be
E[\Delta X_t]=\mu_{pd} \Delta, \,\,\operatorname{Var}(\Delta X_{t})= \sigma^2_{pd}\Delta^2+\sigma_{\mu}^2(t \Delta^2+\frac{\Delta^3}{3}) +\sigma^2 \Delta
\ee
Here, we used the following identities:
\be
\operatorname{Var}(\mu_0 \Delta)= \sigma^2_{pd}\Delta^2,  \,\,\,\,\, \operatorname{Var}\left(\sigma\,(W_{t+\Delta}-W_{t})\right)=\sigma^2 \Delta
\ee
and
\be
\operatorname{Var} \left(\int_t^{t+\Delta} B_s\,ds \right)=\int_t^{t+\Delta} \int_t^{t+\Delta} \operatorname{Cov} (B_u,B_v)du dv=\int_t^{t+\Delta} \int_t^{t+\Delta} \operatorname{min} (u,v)du dv=t \Delta^2+\frac{\Delta^3}{3} 
\ee
Since the drift is stochastic, the increments are non-stationary when $\sigma_{\mu}>0$.

\subsection*{Arithmetic Brownian motion with uncertain drift: discrete-time model}
Let us now approximate the model on a discrete time grid with time step \(\Delta t\) (so that \(t_n = n\,\Delta t\) for \(n=0,1,2,\dots\)). The discrete--time version is as follows. The drift process is given by
\be  
\mu_{n+1} = \mu_n + \sigma_\mu\,\sqrt{\Delta t}\,\eta_n, \quad \mu_0 \sim \mathcal{N}(\mu_{pd}, \sigma_{pd}^2),
\ee
where \(\{\eta_n\}\) is an i.i.d. sequence of standard normal random variables. The price evolution is given by
\be 
X_{n+1} = X_n + \mu_n\,\Delta t + \sigma\,\sqrt{\Delta t}\,\epsilon_n
\ee
where \(\{\epsilon_n\}\) is an i.i.d. sequence of standard normals, independent of \(\{\eta_n\}\).
By iteration we obtain:
\be 
X_n = X_0 + \Delta t \sum_{k=0}^{n-1} \mu_k + \sigma\,\sqrt{\Delta t}\sum_{k=0}^{n-1}\epsilon_k.
\ee
Using  $\mu_k = \mu_0 + \sigma_\mu\,\sqrt{\Delta t}\sum_{i=0}^{k-1}\eta_i$, we have
\be 
X_n = X_0 + n\,\Delta t\,\mu_0 + \sigma_\mu\,\Delta t^{3/2} \sum_{k=0}^{n-1}\sum_{i=0}^{k-1}\eta_i + \sigma\,\sqrt{\Delta t}\sum_{k=0}^{n-1}\epsilon_k.
\ee
The mean is given by
\be 
E[X_n] = X_0 + n\,\Delta t\,\mu_{pd}.
\ee
The variance has three independent contributions. The term \(n\,\Delta t\,\mu_0\) has variance \((n\,\Delta t)^2\,\sigma_{pd}^2\). The noise \(\sigma\,\sqrt{\Delta t}\sum_{k=0}^{n-1}\epsilon_k\) contributes variance \(\sigma^2\,n\,\Delta t\). The double sum term contributes variance 
\be \sigma_\mu^2\,\Delta t^3\, \operatorname{Var}\Biggl(\sum_{k=0}^{n-1}\sum_{i=0}^{k-1}\eta_i\Biggr)=  \sigma_\mu^2\,\Delta t^3\, \operatorname{Var}\Biggl(\sum_{i=0}^{n-1}(n-1-i)\eta_i\Biggr)=  \sigma_\mu^2\,\Delta t^3\,\frac{n(n-1)(2n-1)}{6}.\ee
Thus, the total variance is given by
\be
\operatorname{Var}(X_n)=\sigma^2\,n\Delta t +(n\Delta t)^2 \sigma_{pd}^2  + \sigma_\mu^2\,\Delta t^3\,\frac{n(n-1)(2n-1)}{6}.
\ee
Writing \(t=n\Delta t\), we note that as \(\Delta t\to0\) (with \(t\) fixed) the last term converges to \(\sigma_\mu^2\,t^3/3\). Therefore, the discrete--time result converges to the continuous--time one:
\be 
X_n \sim \mathcal{N}\Bigl(X_0 + \mu_{pd}\,t,\; \sigma^2 t + \sigma_{pd}^2 t^2  + \frac{\sigma_\mu^2 t^3}{3}\Bigr),\quad t=n\Delta t.
\ee
\subsection*{Arithmetic Brownian motion with uncertain volatility}
Assuming the variance of arithmetic Brownian motion is distributed as gamma distribution $s^2\sim \Gamma(\frac{\alpha}{2},\frac{2 \sigma^2}{\alpha})$ and price distribution conditioned on $s$ is  $P(\mu, s; X_T)=\mathcal{N}(\mu T,s^2 T; X_T)$, the distribution of the price with uncertain variance $\tilde{P}(X_T)$ is given by
\be
\tilde{P}(X_T) = \int_0^{\infty} d s^2 \, \Gamma(\frac{\alpha}{2}, \frac{2 \sigma^2}{\alpha};s^2) \, P( \mu, s; X_T)=
\ee
$$
\frac{2^{\frac{1-\alpha}{2}} \alpha ^{\frac{1+\alpha }{4}} (\sigma  \sqrt{T})^{-\frac{1+\alpha }{2}} | X_T-T \mu | ^{\frac{\alpha-1}{2}} K_{\frac{\alpha-1 }{2}}\left(\frac{\sqrt{\alpha } | X_T-T \mu | }{\sqrt{T} \sigma }\right)}{\sqrt{\pi } \Gamma \left(\frac{\alpha }{2}\right)}
$$

\subsection*{Geometric Brownian motion with  drift: continuous-time model}
Consider the GBM given by
\be 
\frac{dS_t}{S_t}=\mu_t\,dt+\sigma\,dW_t,
\ee
with the drift process defined as:
\be 
\mu_t = \mu_0 + \sigma_\mu\,B_t,\qquad \mu_0\sim \mathcal{N}(\mu_{pd},\sigma_{pd}^2),
\ee
and \(W_t\) and \(B_t\) being independent standard Brownian motions. The solution to the GBM SDE is given by
\be 
S_t = S_0\exp\Biggl\{ \int_0^t \mu_s\,ds -\frac{1}{2}\sigma^2 t + \sigma\,W_t\Biggr\}.
\ee
Since $ \int_0^t \mu_s\,ds = \mu_0t + \sigma_\mu\int_0^tB_s\,ds,$
we have
\be 
\ln\frac{S_t}{S_0} = \mu_0t + \sigma_\mu\int_0^tB_s\,ds -\frac{1}{2}\sigma^2t + \sigma\,W_t.
\ee
Given that the three random components are independent, we have
\be 
\mu_0t \sim \mathcal{N}(\mu_{pd} t,\; t^2 \sigma_{pd}^2),\,\, \sigma_\mu\int_0^tB_s\,ds \sim \mathcal{N}\Bigl(0,\; \frac{\sigma_\mu^2 t^3}{3}\Bigr), \,\, \sigma\,W_t \sim \mathcal{N}(0,\; \sigma^2 t).
\ee    
Thus, the log-price is normally distributed:
\be 
\ln\frac{S_t}{S_0} \sim \mathcal{N}\Bigl(\mu_{pd} t-\frac{1}{2}\sigma^2t,\; \sigma^2 t+ \sigma_{pd}^2 t^2+\frac{\sigma_\mu^2t^3}{3}\Bigr).
\ee
Following a derivation similar to that for the increment in ABM, the GBM log-return over \( \Delta \) is given by
\be 
\Delta S_t=\ln \frac{S_{t+\Delta}}{S_t}\sim \mathcal{N} \left((\mu_{pd}-\frac{\sigma^2}{2})\Delta,\sigma^2 \Delta+\sigma^2_{pd}\Delta^2+\sigma_{\mu}^2(t \Delta^2+\frac{\Delta^3}{3})  \right)
\ee 

\subsection*{Geometric Brownian motion with uncertain drift: discrete-time model}
Let \(t_n = n\Delta t\). In discrete-time approximations the drift process is given by
\[
\mu_{n+1} = \mu_n + \sigma_\mu\sqrt{\Delta t}\,\eta_n,\qquad \mu_0\sim \mathcal{N}(\mu_{pd},\sigma_{pd}^2),
\]
with \(\eta_n\sim N(0,1)\) i.i.d. The GBM price process is given by
\[
S_{n+1} = S_n \exp\Bigl\{\mu_n\,\Delta t -\frac{1}{2}\sigma^2\Delta t + \sigma\sqrt{\Delta t}\,\epsilon_n\Bigr\},
\]
with \(\epsilon_n\sim N(0,1)\) i.i.d., independent of \(\eta_n\).
Taking logarithms and iterating, we have
\[
\ln S_n = \ln S_0 + \sum_{k=0}^{n-1}\left[\mu_k\,\Delta t -\frac{1}{2}\sigma^2\Delta t + \sigma\sqrt{\Delta t}\,\epsilon_k\right].
\]
Substituting the representation for $\mu_k = \mu_0 + \sigma_\mu\sqrt{\Delta t}\sum_{i=0}^{k-1}\eta_i,$
we obtain
\[
\ln S_n = \ln S_0 + n\Delta t\,\mu_0 + \sigma_\mu\,\Delta t^{3/2}\sum_{k=0}^{n-1}\sum_{i=0}^{k-1}\eta_i -\frac{1}{2}\sigma^2n\Delta t + \sigma\sqrt{\Delta t}\sum_{k=0}^{n-1}\epsilon_k.
\]
The mean is given by
\be 
E[\ln S_n] = \ln S_0 + n\Delta t\,\mu_{pd} - \frac{1}{2}\sigma^2 n\Delta t.
\ee
The variance is the sum of three components. The term \(n\Delta t\,\mu_0\) has variance \((n\Delta t)^2 \sigma_{pd}^2\). The term \(\sigma\sqrt{\Delta t}\sum_{k=0}^{n-1}\epsilon_k\) has variance \(\sigma^2\,n\Delta t\). The double sum term has variance
$\sigma_\mu^2\,\Delta t^3\,\frac{n(n-1)(2n-1)}{6}.$
Thus,
\[
\operatorname{Var}(\ln S_n) = (n\Delta t)^2 \sigma_{pd}^2 + \sigma^2\,n\Delta t + \sigma_\mu^2\,\Delta t^3\,\frac{n(n-1)(2n-1)}{6}.
\]
Expressing in terms of \(t = n\Delta t\) and noting that
\[
\Delta t^3\,\frac{n(n-1)(2n-1)}{6} \to \frac{t^3}{3} \quad \text{as } \Delta t \to 0,
\]
we recover the continuous--time variance $\sigma^2 t +\sigma_{pd}^2  t^2 +  \frac{\sigma_\mu^2t^3}{3}.$
The log--price is given by
\be 
\ln \frac{S_n}{S_0} \sim \mathcal{N}\Bigl(\mu_{pd} t -\frac{1}{2}\sigma^2t,\; \sigma^2t+\sigma_{pd}^2 t^2 +\frac{\sigma_\mu^2t^3}{3}\Bigr),\quad t=n\Delta t,
\ee
\(S_n\) is lognormally distributed and the discrete model converges to the continuous-time model as \(\Delta t\to 0\).

Thus, we have derived detailed distributional results for both an arithmetic Brownian motion and a geometric Brownian motion with an uncertain, stochastically evolving drift whose initial value is random. In summary, we obtained:
\begin{align}
\text{Arithmetic BM:}\quad X_t-X_0 &\sim \mathcal{N}\Bigl(\mu_{pd} t,\; \sigma^2 t +\sigma_{pd}^2 t^2  +  \frac{\sigma_\mu^2t^3}{3}\Bigr),\\[1mm]
\text{Geometric BM:}\quad \ln\frac{S_t}{S_0} &\sim \mathcal{N}\Bigl(\mu_{pd} t-\frac{1}{2}\sigma^2t,\; \sigma^2 t+\sigma_{pd}^2 t^2 +\frac{\sigma_\mu^2t^3}{3}\Bigr).
\end{align}
The discrete--time formulations were shown to converge to these continuous-time results in the limit as \(\Delta t\to0\).

\subsection*{Geometric Brownian motion with uncertain volatility}
Assume the variance parameter of the GBM is distributed as a gamma distribution $s^2 \sim \Gamma\left(\frac{\alpha}{2}, \frac{2 \sigma^2}{\alpha}\right)$ and the price distribution conditioned on $s$ is $P(X_T) \sim LN(\bar \mu_{\ln}, \bar \sigma_{\ln}; X_T)$, where $\bar \mu_{\ln} = (\mu - \frac{1}{2} s^2) T$ and $\bar \sigma^2_{\ln} = T s^2$. The distribution of the price with uncertain volatility $\tilde{P}(X_T)$ is given by
\be
\tilde{P}(X_T) = \int_0^{\infty} d s^2 \, \Gamma(\frac{\alpha}{2}, \frac{2 \sigma^2}{\alpha};s^2) \, P(\bar \mu_{ln}, \bar \sigma_{ln}; X_T)=
\label{eq:gmb_sigma}   
\ee
$$
\frac{\alpha ^{\alpha /2} e^{\mu T/2} \left(4 \alpha +\sigma ^2 T\right)^{\frac{1-\alpha}{4}}  | T \mu -\log (X_T)| ^{\frac{\alpha -1}{2}} K_{\frac{\alpha -1}{2}}\left(\frac{\sqrt{4 \alpha+\sigma ^2 T } | T \mu -\log (X_T)| }{2 \sqrt{T} \sigma }\right)}{\sqrt{\pi }   \Gamma
   \left(\frac{\alpha }{2}\right) (\sigma  \sqrt{T})^{\frac{1+\alpha }{2}}  X_T^{\frac{3}{2}}}
$$

\section{Appendix F: Kelly leverage for binary betting}
\subsection*{Binary betting: Deterministic parameters}

In the binary setting, the core of the problem that a gambler faces is as follows: Consider a gambler betting on a biased coin, where the probability $p>0.5$ of getting tails is known. Each bet either doubles the money or results in a total loss. While betting all of one's capital maximizes expected wealth, it is extremely risky because it only succeeds on a single path—winning all $N$ bets. As the number of bets $N$ increases, the likelihood of following this path drops to zero, even though the expected wealth, $E[X_N] = X_0 (2p)^N$, continues to grow. This paradox illustrates the trade-off between maximizing growth and avoiding the risk of ruin.

Instead, betting a fixed fraction $f$ of wealth helps mitigate this problem. The optimality of the Kelly criterion is based on two assumptions: that winnings are reinvested and that the number of bets is large. In practice, the number of bets is finite, and the expected value can fluctuate. We model the cost of excessive fluctuations using the variance of logarithmic utility. In this section, we derive the optimal leverage in a binary setting using the GMV.

Following \cite{Thorp:1992} we have. Consider betting a fixed fraction $f$, where $0 < f < 1$, at each trial.  We have the final wealth $X_N$ after random outcomes $Y_i$ which is a win amount $b$ with probability $p$ and loss amount $a$ with probability $q=1-p$:
\be
X_N = X_0 \prod_{i=1}^N (1 + Y_i f)
\ee
Thus the expected value and variance of $\ln X_N$ are:
\be
E[\ln \frac{X_N}{X_0}] = N \, E[\ln(1 + Y_i f)],\,\, \,\,  \text{Var}[\ln \frac{X_N}{X_0}] = N \, \text{Var}[\ln(1 + Y_i f)]
\ee
Then
\be
E[\ln(1 + Y_i f)] = p \ln(1 + b f) + q \ln(1 - a f) \equiv m
\ee
and
\be
\text{Var}[\ln(1 + Y_i f)] = p[\ln(1 + b f)]^2 + q[\ln(1 - a f)]^2 - m^2= pq \left( \ln \left( \frac{1 + b f}{1 - a f} \right) \right)^2
\ee
We observe that the variance grows as $f^2$, and the standard deviation grows linearly with $f$ for small $f$, representing additional uncertainty.

The optimal fraction $f^*$  which maximizes the expected value of logarithmic utility $E[\ln  \frac{X_N}{X_0}]$ is given by
\be
f^*_{K} = \frac{p}{a} - \frac{q}{b}
\label{eq:kelly_bin}
\ee
For this fraction to be positive, it is required that $p b > q a$. This implies that if $p b \leq q a$, the game is not favorable to play.
In the simplest case where $a = b$ and $q = 1 - p$, we have:
\be
f^*_K = 2p - 1
\ee
The generalized mean-variance solution is given by
\be
f^* = \arg \max_{f} \left( E[\ln \frac{X_N}{X_0}] -  \frac{\lambda}{2} \text{Var}[\ln \frac{X_N}{X_0}] \right)
\ee
where the logarithmic utility function encodes risk preference, and the variance term encodes information about dispersion around the expected value. 

The optimal leverage \( f^* \) is the solution of the following non-linear equation, which can be solved numerically:
\be
b p (a f-1)+a q (b f+1)+\lambda  p q (a+b) \log \left(\frac{b f+1}{1-a f}\right)=0
\label{eq:kelly_f1}
\ee
By expanding the logarithm in Eq.~\ref{eq:kelly_f1} up to the first order in \( f \), we have a linear equation in f and the solution is given by
\be
f^* =\frac{b p-a q}{\lambda  p q (a+b)^2+a b (p+q)}
\ee
We define the Kelly multiplier \( \lambda^* \) as:
\be
\delta = \frac{f^* }{f^*_{\text{K}} } = \frac{a b}{\lambda  p q (a+b)^2+a b (p+q)}
\ee
For $a=b=1$ and $\lambda=1$, we have $\delta=1/(1 + 4 (1 - p) p)$. For $p=0.5$, $\delta=1/2$ thus giving the half-Kelly criterion. Expansion of logarithm in Eq.~\ref{eq:kelly_f1} up to the second order in $f$ gives a quadratic equation which has even better approximation  for $f^*$ in the vicinity $p\approx 0.5$. For arbitrary parameters, Eq.~\ref{eq:kelly_f1} can be solved numerically.

\subsection*{Binary betting: A Bayesian model with probabilistic parameters}
In this section, we derive an objective function for optimal leverage with probabilistic parameters.
We calculate the posterior predictive distribution of the wealth in the process of binary betting with uncertain parameters. The optimal leverage is defined as the solution of the GMV with logarithmic utility of wealth.

Assume that for a new sample of \(N\) independent Bernoulli trials $Z_i$ with success probability \(p\), we have:
\be 
Y_i = 
\begin{cases}
b & \text{if } Z_i=1,\\[1mm]
-a & \text{if } Z_i=0.
\end{cases}
\ee
The product over these \(N\) trials with leverage  \(f>0\) is
$X_N =X_0 \prod_{i=1}^{N} (1+f Y_i) = (1+bf)^K (1-af)^{N-K},$ where $K = \sum_{i=1}^{N} Z_i$ denotes the number of successes and $X_0$ is an initial wealth.
Suppose that, based on previous data $D$ (with \(y_1\) successes in \(n_1\) trials) and using a Beta prior  $\operatorname{Beta}(\alpha,\,\beta)$, the posterior distribution for $p$ is given by 
\be P(p \mid \text{D}) = \operatorname{Beta}(y_1+\alpha,\, n_1-y_1+\beta;p).
\label{eq:pd_p}
\ee
Given the posterior for \(p\), the posterior predictive distribution for the number of successes \(K\) in the new \(N\) trials is given by the Beta-Binomial distribution
\be 
P(K=k \mid D) =BetaBin(N, y_1+\alpha,n_1-y_1+\beta;k)
\ee
for \(k=0,1,\dots,N\).
Since
\be 
X_{N} =X_0 (1+bf)^K (1-af)^{N-K},
\ee
the posterior predictive distribution for \({X_N}\) is given by
\be
P\Big(X_{N}=X_0 (1+bf)^k (1-af)^{N-k} \mid D \Big) = BetaBin(N, y_1+\alpha,n_1-y_1+\beta;k).
\ee
Taking the logarithm of \(X_N\) gives:
\be 
\ln \frac{X_{N}}{X_0} = K\ln(1+bf) + (N-K)\ln(1-af)=
N\,B + c\,K.
\ee
where $B = \ln(1-af)$ and $c = \ln\left(\frac{1+bf}{1-af}\right)$.
For the Beta-Binomial distribution, the posterior predictive mean and variance of \(K\) are given by
\be 
E[K] = N \frac{y_1+\alpha}{n_1+\alpha+\beta},
\ee
\be 
\operatorname{Var}(K) = N\,\frac{(y_1+\alpha)(n_1-y_1+\beta)}{(n_1+\alpha+\beta)^2}\,\frac{n_1+\alpha+\beta+N}{n_1+\alpha+\beta+1}.
\ee
Thus, the expected value of \(\ln \frac{X_{N}}{X_0}\) is given by
\be 
E[\ln \frac{X_{N}}{X_0}] = N\,B + c\,E[K] = N\ln(1-af) + N\,\frac{y_1+\alpha}{n_1+\alpha+\beta}\,\ln\left(\frac{1+bf}{1-af}\right).
\ee

Similarly, since \(NB\) is a constant and \(c\) is a constant, the variance of \(\ln \frac{X_{N}}{X_0}\) is $\operatorname{Var}\left[\ln \frac{X_{N}}{X_0}\right] = c^2\,\operatorname{Var}(K),$
or explicitly:
\be 
\operatorname{Var}\left[\ln \frac{X_{N}}{X_0}\right] = \left[\ln\left(\frac{1+bf}{1-af}\right)\right]^2\,N\,\frac{(y_1+\alpha)(n_1-y_1+\beta)}{(n_1+\alpha+\beta)^2}\,\frac{n_1+\alpha+\beta+N}{n_1+\alpha+\beta+1}.
\ee
The optimal leverage $f^*$ is derived by numerically maximizing the GMV criterion:
$$
f^*=\arg\max_f \, \left(E[\ln  \frac{X_{N}}{X_0}]-\frac{\lambda}{2} \text{Var}[\ln \frac{X_N}{X_0}]\right).
$$

The deterministic limit from the previous section is obtained by assuming complete certainty about $p$, which corresponds to the posterior for $p$ becoming a point mass at $p_0 = \frac{y_1+\alpha}{n_1+\alpha+\beta}$. This is achieved by letting the total pseudo-count $T = n_1+\alpha+\beta$ tend to infinity while keeping $p_0$ fixed. 
In that limit, the variance of posterior distribution of $p$ in Eq.~\ref{eq:pd_p}  vanishes, and the posterior predictive distribution for $K$ becomes a standard Binomial distribution $K \sim \operatorname{Binomial}(N, p_0)$.  For arbitrary parameters, the difference between the deterministic and probabilistic GMV cases is mainly due to the fact that the probabilistic variance is scaled by a factor of $\frac{n_1+\alpha+\beta+N}{n_1+\alpha+\beta+1}$ relative to the deterministic variance, while the expected values are often close.

\section{Appendix G: Kelly criterion with power utility function}
The power utility (CRRA) is defined as follows:
\be 
U_r(x,\gamma) =        
    \begin{cases}
        \frac{x^{1-\gamma}-1}{1-\gamma}, &  \mbox{if } \gamma > 0, \gamma \neq 1 \\
        \ln{x} & \mbox{if } \gamma = 1
    \end{cases}
\ee    
Applying the results from Appendix B to the price distribution $P(X_T)\sim  LN(\mu_{\ln}, \sigma_{\ln}; X_T)$, we have:
\be
E[U(X)]= \int_0^{\infty} LN(\mu_{\ln}, \sigma_{\ln}; x) \cdot  U_{r}(x, \gamma) \, dx=\frac{ e^{\frac{1}{2} (1 - \gamma) \left(2 \mu_{\ln} + (1 - \gamma) \sigma_{\ln}^2\right)}-1}{1 - \gamma}
\ee
\be
Var[U(X)]= \frac{e^{2 \mu_{\ln} - 2 \gamma (\mu_{\ln} + 2 \sigma_{\ln}^2)} \left(e^{2 (1 + \gamma^2) \sigma_{\ln}^2}-e^{(1 + \gamma)^2 \sigma_{\ln}^2}  \right)}{(1 - \gamma)^2}
\ee
where $\mu_{\ln} = \left(\mu - \frac{1}{2} \sigma^2\right) T$ and  $\sigma^2_{\ln} = \sigma^2 T$.
The parameter $\gamma$ can be defined using the CE equation, $U(X_{CE})=E[U(X)]$, given an external input of $\mu$, $\sigma$, and a risk profile defined by $x_{CE}$:
$x_{CE}=e^{\mu_{\ln}+\frac{\sigma_{\ln}^2}{2}(1-\gamma)}$.

Assuming leverage $f$, and allocation between cash with a rate of $r_0$ and a portfolio with drift $\mu_p$ and volatility $\sigma_p$, we have:
\be 
\mu = (1 - f) r_0 + f \mu_r, \quad \sigma = f \sigma_r.
\label{eq:musigma}
\ee
The solution for the MEU $\arg \max_f E[U(X)]=\arg \max_f \left[\mu_{ln}+\frac{(1- \gamma)}{2}\sigma_{ln}^2\right]$ is given by
$f^*=\frac{\mu_r-r_0}{\sigma_r^2 \gamma} $.
For $\gamma=1$, we recover the original Kelly leverage
$ f^*=\frac{\mu_r-r_0}{\sigma_r^2} $. For linear utility $\gamma=0$, there is no upper limit on leverage, as it does not penalize risk. Although $\gamma>1$ can provide an alternative explanation of the fractional Kelly, the GMV with logarithmic utility seems to provide a more natural explanation.
The optimal leverage $f^*$ for the GMV objective can be obtained by solving numerically $\arg \max_f E[U(X)]-\frac{\lambda}{2} \text{Var}[U(X)]$.

\section{Appendix E: Practical considerations} 
This section presents our analysis of the preceding results, highlighting their potential applications and the key lessons that can be drawn.

The Markowitz portfolio construction appears conceptually attractive, but in practice,  involves the inversion of a near-singular covariance matrix, which leads to unstable results. Despite this, it is still widely used, but \emph{always} with constraints such as long-only positions, sector allocations, maximum number of positions, turnover limits, and others. It could be that the real added value comes not from MV optimization per se but from the enforcement of these constraints.

If the returns are Gaussian, then the exponential utility and the Markowitz approach yield identical results. However, the advantage of using utility functions is that they can be applied with arbitrary return distributions, which allows us to study the effects of fat tails or skewness in the returns. Moreover, the use of compound distributions allows for the inclusion of uncertainty in both the expected returns and the covariance matrix.

A consistent model of allocation should include input about uncertainties of future expected return and (co)variance. 
The conditional distribution of population variance for Gaussian returns, given sample variance, follows an inverse-gamma distribution, which is wide and exhibits a fat right tail. Thus, even the estimation error (let alone the forecasting error) is large. 
This renders the notion of the efficient frontier and the standard formulation of portfolio optimization in the form of inequalities
\be
\text{maximize} \quad \boldsymbol{\mu}^T \boldsymbol{w},\,\,\, 
\text{subject to:} \,\,
\boldsymbol{w}^T \Sigma \boldsymbol{w} \leq (\sigma_{\text{target}})^2
\ee
fragile.

Diversification has its own limits. A widely diversified portfolio resembles a stock index. In the US, the average performance of the S\&P 500 with dividends is close to 9 percent annually, but it closely matches the M2 monetary supply growth (or money printed), which is around 8 percent. Thus, relative wealth growth can only be created by concentrated bets or leverage. The big winners are few and hard to predict \cite{Markov:2023}. Leverage can help, but it comes at the expense of increasing the probability of ruin.

\subsection*{Effect of fat tails and skewness on diversification}
The asymmetric Laplace distribution strikes a balance by providing a realistic distribution of returns that includes both the main body and tail behavior, as well as skewness, while still allowing for analytical solutions in terms of elementary functions. 
The closed-form solution with exponential utility and ALD returns $r\sim \text{ALD}(\mu,\sigma,\kappa)$ in the one-dimensional case provides a clear picture of the influence of fat tails and skewness on risk allocation \cite{MarkovA:2023}. We observe that fat tails  saturates the solution for large $\mu$ relative to the MV solution. The MV solution grows linearly with $\mu$, as $w^*_{\text{MV}} = \dfrac{\mu}{a \sigma^2}$, while the ALD solution for large $\mu$ saturates to a function that does not depend on $\mu$:
\be
w^*_{\text{ALD}} = \frac{\sqrt{2}}{a \sigma \kappa} + O\left( \frac{1}{\mu} \right)
\label{eq:univ-ald-large-mu}
\ee
As can be seen, the weights in Eq.~\ref{eq:univ-ald-large-mu} do not grow indefinitely, as in the Markowitz solution $w^*_{\text{MV}}$, but instead saturate and become independent of the signal  strength. This allocation rule can be relevant for quantitative strategies with a high signal-to-noise ratio.

The skewness introduces a threshold on allocations, as only assets with $\mu > \sqrt{2} (\kappa - 1) \sigma$ are worth investing in. The ALD solution for small skewness is given by
\be
w^*_{\text{ALD}} = -\sqrt{2} \frac{(\kappa - 1)}{a \sigma} + \frac{\mu}{a \sigma^2} + O\left( \left( \frac{\mu}{\sigma} \right)^2 \right)
\ee
The effect of skewness cannot be disregarded in some popular strategies such as trend-following strategies (which exhibit positive skewness) and selling options (which exhibits negative skewness).

Although multivariate ALD allows a closed-form solution, it is difficult to calibrate and has a few counter-intuitive properties. Instead, one can generate a fat-tailed distribution from the scale mixture approach by integrating a Gaussian distribution over variances. The drawback, though, is that it is not possible to obtain a skewed distribution in this approach.

The effect of uncertain variance can be modeled using a model with exponential utility, assuming normally distributed returns and uncertain variance distributed according to $ s^2 \sim \Gamma\left(\frac{\alpha}{2}, \frac{2\sigma^2}{\alpha}\right)$ \cite{MarkovB:2023}.
The asymptotic for $\alpha \to \infty$  (small variance uncertainty) is given by
\be
w_{GD}^*=\frac{\mu}{a\sigma^2}-\frac{\sigma^4 \mu^3}{a \alpha \sigma^8}+O(\alpha^{-\frac{3}{2}})  
\label{eq:univ-asympt-1}
\ee 
 
The leading term corresponds to the MV solution. We note that an increase in  $\alpha$ (a decrease in variance uncertainty) increases in the allocation to the risky asset  $w^*$. 

The asymptotic for $\alpha \to 0$  (large variance uncertainty) is given by
\be
w_{GD}^*=\frac{\sqrt{\alpha}}{a\sigma}-\frac{\alpha}{2 a \mu}+O(\alpha^\frac{3}{2}) 
\label{eq:univ-asympt-2}
\ee 
The leading term corresponds to the inverse volatility allocation. 

\subsection*{Worst-case optimization}
An alternative to the expected utility, or average scenario, or mean-variance optimization is a minimax or worst-case optimization. In this scenario, a  modeler comes with an appropriate worst-case scenario for the expected return or for the covariance matrix and optimizes decision variables based on this scenario. The objective of the worst-case optimization is the following: 
\be
\begin{aligned}
& \arg \max_{\bs w} \quad \min_{\bs \mu \in D_{\bs \mu}} \left( \bs w^T \bs \mu \right) - \frac{a}{2} \max_{\bs \Sigma \in D_{\bs \Sigma}} \left( \bs w^T \bs \Sigma \bs w \right) \\
\end{aligned}
\ee
Here a modeler is concerned with a realistic description of the uncertainty domains  $ D_{\bs \mu}$ and $D_{\bs \Sigma}$ while  allowing the optimization problem to be solvable. 
Worst-case optimization is often used anywhere where the probability of ruin is non-zero. 

\subsection*{Worst-case optimization comparison: expected returns}

One of the most common definition of $D^e_{\bs \mu}$ is the elliptical uncertainty set, which is defined as follows:
\be
D^e_{\bs \mu} = \left\{ \bs \mu \, \big| \, ( \bs  \mu - \hat{  \mu})^T  \bs  S_{ \bs \mu}^{-1} ( \bs  \mu -   \hat{\mu}) \leq \delta^2_{ \bs \mu} \right\},
\ee
where the predefined parameters $ \hat{\mu}$, $\delta_{ \bs \mu}$, and $\bs S_{\bs \mu}$ describe the uncertainty set \cite{Lobo:2000}. In case of diagonal $\bs S_{\bs \mu}$, we have a weighted sum of squared errors limited by a constant $\delta_{\bs \mu}^2$ or, in another words, $L_2$ penalty for allocation weights.

An alternative approach is a  minimax portfolio (MM) \cite{MarkovA:2023} where the optimization is done for a portfolio where the maximum allocation is to the worst performing asset assuming that the expected returns follow a multivariate normal distribution with mean $\boldsymbol \mu_0$ and diagonal variance $\boldsymbol \Sigma_0$ such that  $\boldsymbol \mu \sim N(\boldsymbol \mu_0,\boldsymbol \Sigma_0) = \boldsymbol \mu_0+\boldsymbol Y$ and $\boldsymbol Y\sim N(0,\boldsymbol \Sigma_0)$. 
This leads to the infinity norm penalty $\max_i w = |w|_{\infty}$:
\begin{equation}
\boldsymbol w^*=\arg \max_w\left[ \boldsymbol \mu_0^T \boldsymbol w+\min \boldsymbol  Y *\max_i w_i -\frac{a}{2} \boldsymbol w^T \Sigma \boldsymbol w\right], \,\, s.t. \sum_{i=1}^N w_i=1, \, w_i\ge 0
\label{gauss-min-worst-optimal}
\end{equation}    
The solution can be obtained analytically using the Karush–Kuhn–Tucker (KKT) formalism or numerically via a convex optimizer. The solution is a mixture of an equal-weights allocation for the top weights and the MV solution for the smaller weights.

The difference between the two approaches can be understood by examining the difference between the $L_2$ and $L_\infty$ norms. The $L_2$ norm promotes overall weight shrinkage and smoothness of the weight distribution, while the $L_\infty$ norm's primary goal is to limit concentration risk. An advantage of an $L_\infty$ penalty is that it is parameterized by a single parameter and has an analytical solution.

The model in Eq.~\ref{gauss-min-worst-optimal} can be viewed as a toy model of a robust political system, where the concentration of power is a major risk and optimal allocation of power is required. The solution is an equal allocation of power at the top (a management committee) and the mean-variance approach for the bottom allocations (meritocracy). The worst-case scenario occurs when maximum power is given to a politician who makes appealing promises but ultimately delivers the poorest outcomes after the election and upon assuming office.

\subsection*{Worst-case optimization: covariance matrix}

Future covariance matrices can deviate from a historical estimate $\bs \Sigma$. Long-only equity investing has two primary sources of uncertainty. The covariance matrix estimate is noisy due to non-stationarity and limited sample size and it can be captured by the Wishart noise model $\bs S \sim W_N(\alpha,\frac{\bs \Sigma}{\alpha})$, which we discussed above.
The other source of uncertainty is related to regime changes in the correlation structure, conventionally referred to as risk-on/risk-off regimes, where in a risk-off regime, stocks become correlated and move in unison.

To account for worst-case scenarios, a risk-on/risk-off framework can be employed, which utilizes two covariance matrices: one for the normal regime and another for the market under stress. This approach assumes that a modeler assigns a non-zero probability of a market crash or correction occurring during the portfolio's expected holding period. Operating in a discrete space is more practical in this context.

Specifically, the future is modeled as a two-state system: a normal regime with parameters \( \boldsymbol{\mu}_n \) and \( \boldsymbol{\Sigma}_n \), and a stressed regime with parameters \( \boldsymbol{\mu}_s \) and \( \boldsymbol{\Sigma}_s \). Within this framework, the expected value of the utility function is a discrete sum, with probability \( p \) assigned to the normal regime and probability \( 1 - p \) to the stressed regime. Consequently, the expected utility of the two-state system \( E_{2s}[U_a] \) is given by
\be
E_{2s}[U_a] = p \, E[U_a(\boldsymbol{\mu}_n, \boldsymbol{\Sigma}_n)] + (1 - p) \, E[U_a(\boldsymbol{\mu}_s, \boldsymbol{\Sigma}_s)]
\label{eq:two-state-1}
\ee
Taking the logarithm of Eq.~\ref{eq:two-state-1}, we derive the following optimization problem:
\be
\boldsymbol{w}^* = \arg \min_{\boldsymbol{w}} \left[ \log\left( \exp(u_n) + \exp(u_s) \right) \right]
\label{eq:two-state-2}
\ee
where
\be
u_n = \log(p) + \frac{a^2}{2} \boldsymbol{w}^T \boldsymbol{\Sigma}_n \boldsymbol{w} - a \boldsymbol{\mu}_n^T \boldsymbol{w},\,\,\,
u_s = \log(1 - p) + \frac{a^2}{2} \boldsymbol{w}^T \boldsymbol{\Sigma}_s \boldsymbol{w} - a \boldsymbol{\mu}_s^T \boldsymbol{w}.
\ee
The covariance matrix of the stressed regime $u_s$ can be taken as an equicorrelation matrix:
\be
\bs \Sigma_s = \sigma_s^2 \left[(1 - \rho_s)\bs I + \rho_s \bs 1\bs 1^T \right]
\ee
The risk for an equal-weight allocation $w = \frac{1}{N}$ is given by 
\be
R = \bs w^T \bs \Sigma_s \bs w = \frac{\sigma_s^2}{N} \left(1 + \rho_s (N - 1)\right)
\ee
An important fact about this approximation is its finite residual risk, which cannot be removed by diversification. Even if the number of assets becomes very large $N \to \infty$, we have $R = \sigma_s^2 \rho_s$. This risk must be hedged separately. 

\subsection*{Comparison of minimax and risk parity portfolios}
Naive minimum variance portfolio allocation often leads to uneven allocations due to the inversion of the ill-conditioned covariance matrix. Given that the sum of weights is normalized, there are two primary ways to address over-allocation: increasing the contribution of small weights and mitigating large weights. The log penalty used in risk parity (RP)  penalizes  small allocations, while the MM portfolio, which employs the infinity norm, penalizes large allocations, making the portfolio weights more uniform. Thus, the MM portfolio presents an alternative to RP for dealing with over-allocation.

RP is an approach to portfolio management that focuses on allocating risk rather than  capital. The RP portfolio seeks to ensure that each asset contributes equally to the portfolio's overall volatility.
Spinu \cite{Spinu:2013} demonstrated that the RP allocation problem can be formulated as the following convex optimization problem:
\be
\begin{aligned}
\boldsymbol{w}^* = \arg \min_{\boldsymbol{w}} \left[\frac{1}{2} \boldsymbol{w}^T \bs \Sigma \boldsymbol{w} - b \sum_{i=1}^N \log(w_i) \right],\,\,\text{subject to} \quad \sum_{i=1}^N w_i = 1, \quad w_i \geq 0, \quad b > 0
\end{aligned}
\ee
The logarithmic penalty is significant for small weights, thereby encouraging a more balanced allocation by increasing the allocation to smaller positions. The MM portfolio is defined as follows:
\begin{equation}
\boldsymbol{w}^* = \arg \min_{\boldsymbol{w}} \left[ \frac{1}{2} \boldsymbol{w}^T \bs \Sigma \boldsymbol{w} + c \max_i w_i \right], \quad \text{subject to} \quad \sum_{i=1}^N w_i = 1, \quad w_i \geq 0, \quad c > 0
\label{mm-min-worst}
\end{equation}
As both terms in the objective function are convex, the optimization problem is also convex, ensuring the existence of a unique global optimum. The infinity norm penalty \(|w|_{\infty}= \max_i w_i \) imposes a large cost on large weights, enforcing equal allocation among large positions and minimizing variance allocation for smaller weights. In principle, the two penalties can be combined to achieve both objectives.

\begin{thebibliography}{14}
\providecommand{\natexlab}[1]{#1}
\providecommand{\url}[1]{\texttt{#1}}
\expandafter\ifx\csname urlstyle\endcsname\relax
  \providecommand{\doi}[1]{doi: #1}\else
  \providecommand{\doi}{doi: \begingroup \urlstyle{rm}\Url}\fi



\bibitem[Markowitz(1952)]{Markowitz:1952}
Harry Markowitz.
\newblock Portfolio selection.
\newblock \emph{The Journal of Finance}, Vol.7, No.\penalty0 1:\penalty0 77, 1952.


\bibitem[Arrow(1966)]{Arrow:1966}
Kenneth J. Arrow and Yrj{\"o} Jahnssonin s{\"a}{\"a}ti{\"o}.
\newblock Aspects of the theory of risk-bearing.
\newblock \emph{Economica}, Vol. 33, \penalty0 251, 1966.

\bibitem[Pratt(1964)]{Pratt:1964}
John~W. Pratt.
\newblock Risk aversion in the small and in the large.
\newblock \emph{Econometrica}, Vol. 32, No.\penalty0 1:\penalty0 122, 1964.

\bibitem[Kelly(1956)]{Kelly:1956}
John Kelly.
\newblock A new interpretation of information rate.
\newblock \emph{IRE Transactions on Information Theory}, Vol. 2, No.\penalty0 3:185, 1956


%
%
\bibitem[Markov and Markov(2023A)]{MarkovA:2023}
Maxime Markov and Vladimir Markov.
\newblock Portfolio optimization rules beyond the mean-variance approach.
\newblock \emph{arXiv:2305.10403}, 2023.
\newblock URL \url{https://doi.org/10.48550/arXiv.2305.08530}.

\bibitem[Markov and Markov(2023B)]{MarkovB:2023}
Maxime Markov and Vladimir Markov.
\newblock Optimal portfolio allocation with uncertain covariance matrix
\newblock \emph{arXiv:2311.07478}, 2023.
\newblock URL \url{https://doi.org/10.48550/arXiv.2311.07478}.

\bibitem[Birge and Chavez-Bedoya(2020)]{Birge:2020}
John~R. Birge and Luis Chavez-Bedoya.
\newblock Portfolio optimization under the generalized hyperbolic distribution:
  Optimal allocation, performance and tail behavior.
\newblock \emph{Quantitative Finance}, Vol. 21, No.\penalty0 2:\penalty0 199,
  2020.

\bibitem[Luxenberg and Boyd(2023)]{Boyd:2022}
Eric Luxenberg and Stephen Boyd.
\newblock Portfolio construction with Gaussian mixture returns and exponential
  utility via convex optimization.
\newblock \emph{Optimization and Engineering}, 2024.


\bibitem[Giller(2004)]{Giller:2004}
Graham L. Giller.
\newblock Frictionless asset allocation with elliptically symmetric distributions of returns.
\newblock URL: \url{https://ssrn.com/abstract=1300671}.

\bibitem[Bodnar(2020)]{Bodnar:2020}
D. Bauder, T. Bodnar, N. Parolya, and   W. Schmid.
\newblock Bayesian mean–variance analysis: optimal portfolio selection under parameter uncertainty.
\newblock \emph{Quantitative Finance}, Vol. 21, No.\penalty0 2:\penalty0 221, 2020.


\bibitem[Cochrane(2011)]{Cochrane:2011}
John H. Cochrane.
\newblock Discount Rates.
\newblock \emph{NBER Working Paper}, No.\penalty0 16972, 2011.
\newblock URL: \url{http://www.nber.org/papers/w16972}.

\bibitem[Durbin(2012)]{DurbinKoopman:2012}
James Durbin and Siem Jan Koopman.
\newblock \emph{Time Series Analysis by State Space Methods}.
\newblock 2nd edn. Oxford University Press, 2012.

\bibitem[Chui and Chen(2017)]{ChuiChen:2017}
Charles K. Chui and Guanrong Chen.
\newblock \emph{Kalman Filtering: with Real-Time Applications}.
\newblock Springer, 2017.
 


%



\bibitem[Rotando and Thorp(1992)]{Thorp:1992}
Louis M. Rotando and Edward O. Thorp.
\newblock The Kelly criterion and the stock market.
\newblock \emph{The American Mathematical Monthly}, Vol.99, No.\penalty0 10:\penalty0
  922, December 1992.  
%

%
%


%


\bibitem[Spinu(2013)]{Spinu:2013}
Florin Spinu.
\newblock An algorithm for computing risk parity weights.
\newblock \emph{Econometric Modeling: Capital Markets - Portfolio Theory eJournal}, 2013.
\newblock URL \url{https://ssrn.com/abstract=2297383}.


\bibitem[Rego et al.(2021)]{Rego:2021}
B.~V. Rego, D.~Weiss, M.~R. Bersi, and J.~D. Humphrey.
\newblock Uncertainty quantification in subject-specific estimation of local vessel mechanical properties.
\newblock \emph{International Journal for Numerical Methods in Biomedical Engineering}, Vol.37, No. \penalty0  12, 2021. 

\bibitem[Lobo and Boyd (2000)]{Lobo:2000}
Miguel S. Lobo and Stephen Boyd. 
\newblock The worst-case risk of a portfolio.
\newblock Technical report.
\newblock URL: \url{https://web.stanford.edu/~boyd/papers/pdf/risk_bnd.pdf}.

%
%


\bibitem[Markov and Markov(2023)]{Markov:2023}
Maxime Markov and Vladimir Markov.
\newblock The impact of big winners on passive and active equity investment
  strategies, 2022.
\newblock URL: \url{https://doi.org/10.48550/arXiv.2210.09302}.

\end{thebibliography}
\end{document}